\documentclass[%
	pre,
	twocolumn,
	superscriptaddress,
	footinbib,
	showpacs,
	shwokeys,
	twoside,
	preprintnumbers,
  ]{revtex4}[2001/08/03]
\usepackage{amsmath}
\usepackage{amssymb}
\usepackage{mathrsfs}
\usepackage{graphicx}
\usepackage{color}
\usepackage{ifpdf}
\ifpdf
\DeclareGraphicsExtensions{.mps,.ps,.eps}
\usepackage{hyperref}
\hypersetup{%
	pdftitle={Spherical Vesicles Distorted by a Grafted Latex Bead: An Exact Solution},
	pdfauthor={\firstname{J\'er\^ome} \surname{Benoit} and \firstname{Avadh} \surname{Saxena}},
	pdfsubject={PACS: 87.16.Dg, 11.10.Lm, 02.40.-k},
	pdfkeywords={exact solution; bending elasticity; elastic compatibility; Bogomol'nyi decomposition},
	pdfpagemode=UseNone,
	pdfstartpage=1,
	pdfhighlight=/O,
	pdfview=FitH,
	pdfstartview=FitH,
	linkcolor=blue,
	citecolor=blue,
	urlcolor=blue,
  }
\IfFileExists{thumbpdf.sty}{\usepackage{thumbpdf}}{}
\providecommand{\eprint}[2][]{\href{http://xxx.lanl.gov/abs/#2}{\texttt{#2}}}%
\else
\DeclareGraphicsExtensions{.mps,.ps,.eps}
\fi
\IfFileExists{datetime.sty}{%
	\usepackage{datetime}
	\providecommand{\now}{\xxivtime\space\today}
	}{%
	\providecommand{\now}{\the\time,\space\today}
	}

\begin{document}
\title[Spherical Vesicles Distorted by a Grafted Latex Bead]{%
  Spherical Vesicles Distorted by a Grafted Latex Bead: An Exact Solution}
\author{\firstname{J\'er\^ome} \surname{Benoit}}%
\email{jgmbenoit@mailsnare.net}
\affiliation{%
	Graduate Institute of Biophysics and Center for Complex Systems,
	National Central University,
	300, Jhongda Road,
	Jhongli City,
	Taoyuan, Taiwan 320,
	Taiwan%
  }
\affiliation{%
	Physics Department,
	University of Crete and Foundation for Research and Technology-Hellas,
	P. O. Box 2208,
	GR-71003 Heraklion,
	Crete,
	Greece%
  }
\author{\firstname{Avadh} \surname{Saxena}}%
\email{avadh@lanl.gov}
\affiliation{%
	Theoretical Division and Center for Nonlinear Studies,
	Los Alamos National Laboratory,
	Los Alamos,
	New Mexico 87545,
	USA%
  }
\date{\eprint{cond-mat/0404250}}
\begin{abstract}
We present an exact solution to the problem of the global shape description 
of a spherical vesicle distorted by a grafted latex bead.
This solution is derived by treating the nonlinearity in bending elasticity
through the (topological) Bogomol'nyi decomposition technique
and elastic compatibility.
We recover the ``hat-model'' approximation in the limit of a small latex bead and
find that the region antipodal to the grafted latex bead flattens.
We also derive the appropriate shape equation using the variational principle 
and relevant constraints.
\end{abstract}
\maketitle

\section{Introduction}
\label{sec/introduction}
Encapsulation, binding and adsorption of particles
onto a membrane play an important role
in biological processes.
A prominent issue in the context of vesicles is
the understanding of long-range phenomena connected to
the inclusion of mesoscopic particles in their membrane
\cite{ISF,Goulian1993,Dommersnes1998,Koltover1999,Gozdz2007}.
The difference in scale between the vesicle's membrane thickness
and the long-range limit (\textit{e.g.}, vesicle size) allows the vesicle
to be described as an embedded surface.
The description of a vesicle's membrane behavior in the long-range limit
is currently founded on the concept of bending elasticity
\cite{Canham,HelfrichZN1,HelfrichZN2,CFMV}.
Recently the Bogomol'nyi decomposition technique \cite{BelavinPolyakov,Bogomolnyi}
has emerged as a promising theoretical framework to study long-range,
nonlinear elastic phenomena in soft-condensed matter
\cite{SantangeloKamien2005,SantangeloKamien2003,BDVAG,SDBTV}.

Exact solutions to nonlinear models offer novel insights into physical 
systems not at first unveiled by approximate approaches.
Therefore our goal here is to derive and discuss an exact solution,
which is motivated by the experimental observation of a spherical vesicle
distorted by a grafted latex bead \cite{Koltover1999},
within the Bogomol'nyi framework combined with elastic compatibility
and a global constraint.
To this end, in the next section we introduce the bending Hamiltonian,
describe the fundamental theorem of surface theory,
introduce the Bogomol'nyi decomposition technique,
and impose a conformally invariant global constraint
meant to mimic the total mean curvature, area, and/or volume constraints.
In {Sec.}~\ref{sec/sphericalshapeequation}
we invoke the variational principle to derive the relevant shape equation.
In {Sec.}~\ref{sec/localversusglobal}
we derive an exact solution for the deformation of the vesicle
in the presence of a grafted latex bead at its north pole.
This solution is obtained by noticing that
the polar angle of the outward surface normal is a kink soliton 
solution of the double sine-Gordon equation.
We also discuss the resultant geometric frustration.
We summarize our main findings
in {Sec.}~\ref{sec/conclusion}.
Finally,
the details of the metric and shape tensors in isometric azimuthal coordinates
are given in Appendix~\ref{appendix/A}
whereas the shape equation is discussed in Appendix~\ref{appendix/B}.

\section{Model}
\label{sec/model}
\subsection{Preliminaries}
In order to take advantage of the Bogomol'nyi technique,
we should describe the vesicle shape within a covariant field theory
\cite{BDVAG,SDBTV}.
According to the fundamental theorem of surface theory
\cite{Struik,GPI},
any embedded surface is perfectly represented by
a pair of symmetric second-rank tensors
coupled to each other by integrability conditions:
a prescribed metric tensor
${\tens{g}}_{ij}$ coupled to a prescribed shape tensor
${\tens{b}}_{ij}$
\textit{via} elastic compatibility conditions.
Within this framework,
the bending Hamiltonian suggested by Canham
\cite{Canham}
reads:
\begin{equation}
\label{BP/Hb/functional/def}
\stHmBnd\left[\stSurface\right]%
	\equiv%
  \tfrac{1}{2}\stKb\!%
	{\int_{\stSurface}}\!%
  \stdS\;%
  {\tens{b}}_{ij}{\tens{b}}^{ij}%
  ,
\end{equation}
which depends on the vesicle shape $\stSurface$ through
the prescribed pair $({\tens{g}}_{ij},{\tens{b}}_{ij})$
and on the phenomenological bending rigidity $\stKb$.
We have adopted the Einstein summation convention
\cite{EinsteinSummationConvention},
and used customary notation:
the integral runs over the surface manifold $\stSurface$
with surface element $\stdS=\std{x^2}\sqrt{\left|g\right|}$,
where $\left|g\right|$ represents the determinant $\det({\tens{g}}_{ij})$
and $x$ the set of arbitrary intrinsic coordinates.
Since the roundness of the grafted latex bead imposes
axisymmetric deformations,
we focus on vesicles of revolution.
In a recent work \cite{SDBTV}, we have shown that,
for surfaces of revolution,
the isometric azimuthal coordinates \mbox{$(u,\varphi)$}
are a relevant choice, which not only drastically simplifies
formulas related to the fundamental theorem of surface theory but 
also elucidates the application of the Bogomol'nyi technique to the 
bending Hamiltonian (\ref{BP/Hb/functional/def}).

\subsection{Fundamental Theorem of Surface Theory}
Working with abstract manifolds
described by a parametrized metric tensor ${\tens{g}}_{ij}$
is common in general relativity
\cite{TMTBH}:
when the abstract manifold is embedded,
the metric tensor ${\tens{g}}_{ij}$ may be coupled
with a shape tensor ${\tens{b}}_{ij}$.
For embedded bidimensional abstract manifolds,
the coupling occurs through integrability conditions
(the Gauss-Codazzi equations)
and a system of differential equations to integrate
(the Gauss-Weingarten equations under appropriate additional conditions):
this is the substance on which the statement and the proof of the
fundamental theorem of surface theory are based
\cite{Struik,GPI}.
Following the fundamental theorem of surface theory allows us to claim that
any pair of  diagonal second-rank tensors \mbox{$({\tens{g}}_{ij},{\tens{b}}_{ij})$},
which in isometric azimuthal coordinates \mbox{$(u,\varphi)$},
takes the form
\begin{gather}
\label{BP/SVR/metric}
	{\tens{g}}_{uu}=%
		{\tens{g}}_{\varphi\varphi}=\stE^{2\stWeylScalingExp(u)}%
	,\\
\label{BP/SVR/shape}
	{\tens{b}}_{uu}=%
		-\stE^{\stWeylScalingExp(u)}%
		\partial_{u}\stNPolarAngle(u)%
	\quad\text{and}\quad%
	{\tens{b}}_{\varphi\varphi}=%
		-\stE^{\stWeylScalingExp(u)}%
		\sin\stNPolarAngle(u)%
	,
\end{gather}
and obeys the elastic compatibility condition
\begin{equation}
\label{BP/SVR/intcond}
	\partial_{u}\stWeylScalingExp(u)%
	=%
		\cos\stNPolarAngle(u)%
	,
\end{equation}
where
the local Weyl gauge field $\stWeylScalingExp$
and the polar angle $\stNPolarAngle$ of the outward surface normal
are sufficiently  differentiable functions of $u$,
corresponds to a unique axisymmetric surface $\stSurfaceOfRevolution$
---
modulo its position in space.
Detailed in Appendix~\ref{appendix/A},
the demonstration consists in showing
that the associated Gauss-Weingarten equations
under appropriate additional conditions determine
a unique surface of revolution
modulo an arbitrary rigid motion.
In simple terms,
straightforward successive integrations
[computational chain (\ref{appendix/reconstruction/X/step/first/preliminary})--(\ref{appendix/reconstruction/X/step/fourth})] 
reduce the set of differential equations to integrate
[(\ref{appendix/converse/sode/first})--(\ref{appendix/converse/sode/last})] 
to the pair of equations:
\begin{gather}
\label{BP/SVR/GW/radius}
	\phantom{\partial_{u}}{r(u)}=%
		\stE^{\stWeylScalingExp(u)}%
	,\\
\label{BP/SVR/GW/height}
	\partial_{u}{z(u)}=%
		-%
		\stE^{\stWeylScalingExp(u)}%
		\sin\stNPolarAngle(u)%
	,
\end{gather}
where $r(u)$ and $z(u)$ are, respectively,
the radius and the height
of the axisymmetric surface
in cylindrical parametrization.
Formula (\ref{BP/SVR/GW/radius})
is a reparametrization that gives
\textit{a posteriori}
a second interpretation to $\stWeylScalingExp$.
In short,
for abstract surfaces of revolution
(\ref{BP/SVR/metric})--(\ref{BP/SVR/shape}),
the integrability conditions reduce
to the single condition (\ref{BP/SVR/intcond})
while the system of differential equations to be integrated simplifies 
to the single equation (\ref{BP/SVR/GW/height}).
Therefrom,
the bending Hamiltonian (\ref{BP/Hb/functional/def})
can be written as
\begin{equation}
\label{BP/SVR/Hb/canonic/Hamiltonian}
\stHmBnd\left[\stSurfaceOfRevolution\right]%
	=%
	\pi\stKb%
	{\int_{\stProfile}}\!%
	\std{u}\;%
	\left[%
		\bigl(%
			{\partial_{u}\stNPolarAngle(u)}%
		\bigr)^{2}%
		+%
		{\sin^{2}\stNPolarAngle(u)}%
	\right]%
	,
\end{equation}
where the integral runs along the profile $\stProfile$
of the axisymmetric surface $\stSurfaceOfRevolution$.

\begin{figure}[t]
	\begin{center}
		\includegraphics[width=0.85\linewidth]{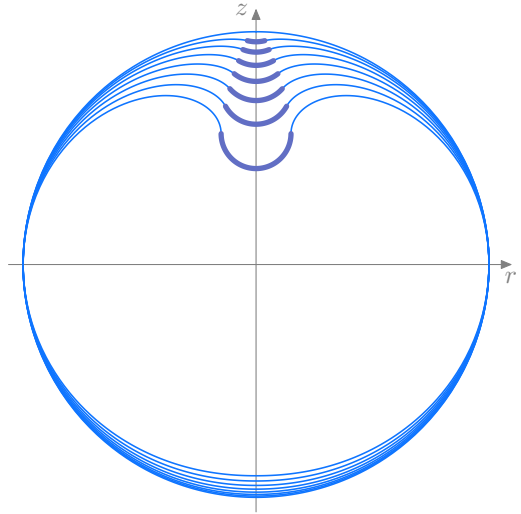}%
	\end{center}
	\caption{
		Cross-sectional profiles
		of a spherical vesicle of revolution
		distorted by a round latex bead
		grafted at its north pole
		with respect to different encapsulated radii $\stBeadEncapsulatedRadius$.
		The bold arcs indicate
		the polar cap imposed by the bead.
		Dimensionless latex bead radius $\stBeadRadius\!=\!0.15$;
		relative encapsulated radii ${\stBeadEncapsulatedRadius}/{\stBeadRadius}$
		from outside to inside:
		$0$,
		$\frac{1}{4}$,
		$\frac{3}{8}$,
		$\frac{1}{2}$,
		$\frac{5}{8}$,
		$\frac{3}{4}$,
		$\frac{7}{8}$,
		and
		$1$.
		}%
\label{fig/GlobalCrossSections}
\end{figure}

\subsection{Bogomol'nyi Decomposition}
The Bogomol'nyi technique allows us to resolve
the Hamiltonian
(\ref{BP/Hb/functional/def})
into a perfect square Hamiltonian and a topological bound
\cite{BDVAG,SDBTV}: 
for spherical surfaces of revolution $\stSurfaceOfRevolution_{0}$,
the decomposition reads
\cite{SDBTV}
\begin{equation}
\label{BP/SVR/Hb/Bogo/decomposition/Hamiltonian}
\stHmBnd\left[\stSurfaceOfRevolution_{0}\right]%
	=%
	\pi\stKb%
	{\int_{\stProfile}}\!%
	\std{u}\;%
	\bigl[
		\partial_{u}\stNPolarAngle(u)%
		-%
		\sin\stNPolarAngle(u)%
	\bigr]^2%
	+%
	4\pi\stKb%
	,
\end{equation}
which readily saturates
the bound
when
the polar angle
$\stNPolarAngle$ satisfies
the first-order nonlinear differential equation
\cite{SDBTV}:
\begin{equation}
\label{BP/SVR/Hb/saturation/equation}
	\partial_{u}\stNPolarAngle(u)=\sin\stNPolarAngle(u)%
	.
\end{equation}
The centered solution of (\ref{BP/SVR/Hb/saturation/equation}) is the
aperiodic
sine-Gordon kink:
\begin{equation}
\label{BP/SVR/Hb/saturation/SG/arctan}
	\stNPolarAngle(u)=%
		2\arctan{\stE^{u}}%
	.
\end{equation}
As expected \cite{BDVAG,SDBTV},
the axisymmetric surface obtained
by integrating the related Gauss-Weingarten equations
under appropriate conditions
(Appendix~\ref{appendix/A})
is the round sphere,
in cylindrical parametrization,
\begin{equation}
\label{BP/SVR/Hb/saturation/surface}
	r(u)=%
		\;
		\sech{u}%
	,%
	\qquad%
	z(u)=%
		-\tanh{u}%
	,
\end{equation}
the isometric coordinate $u$ running from $-\infty$ to $+\infty$.
This remarkable result has encouraged us \cite{SDBTV} to envision 
\textit{deformed} spherical vesicles of revolution as frustrated or unsaturated
sine-Gordon kinks \cite{SDBTV,TSGF,HSETS}.

\subsection{System of Investigation}
Now let us assume a ``bare'' spherical vesicle of revolution distorted 
by a round latex bead chemically grafted at its north pole
\cite{Koltover1999}.
By bare, we mean that only the bending energy 
(\ref{BP/Hb/functional/def}) is considered.
Furthermore,
to mimic the total mean curvature, area and/or volume constraints
\cite{CFMV,STVPDSCBCM},
we impose the following covariantlike global constraint:
\begin{equation}
\label{BP/GC/functional/def}
	{\int_{\stSurface}}\!%
  \stdS\;%
	{\left|g\right|}^{-{\frac{1}{2}}}%
	=%
		\stGCconst%
 	.
\end{equation}
As the metric determinant $\left|g\right|$ is not a (covariant) scalar,
the global constraint (\ref{BP/GC/functional/def}) is not covariant:
in other words,
because it depends on the choice of coordinates,
the global constraint (\ref{BP/GC/functional/def})
is apparently unphysical.
On the other hand, the particular choice of the 
exponent applied to the metric determinant $\left|g\right|$
renders the global constraint (\ref{BP/GC/functional/def})
locally invariant under conformal transformations:
hence, contrary to the (physical) area and/or volume constraints,
the (unphysical) global constraint (\ref{BP/GC/functional/def}) is conformally 
invariant just like the bending Hamiltonian (\ref{BP/Hb/functional/def}).
Also the intuitive motivation behind the choice of (\ref{BP/GC/functional/def})
as a global constraint appears naturally in isometric azimuthal 
coordinates \mbox{$(u,\varphi)$} as then we have
\begin{equation}
\label{BP/GC/functional/explicit}
	2\pi%
	{\int_{\stProfile}}\!\!%
	\std{u}\;%
	=%
		\stGCconst%
	.
\end{equation}
Thus, ultimately our system is governed by the Hamiltonian
(\ref{BP/SVR/Hb/canonic/Hamiltonian})
[or (\ref{BP/SVR/Hb/Bogo/decomposition/Hamiltonian})] 
which must be minimized subject
to the elastic compatibility condition (\ref{BP/SVR/intcond}),
to the geometrical confinement (\ref{BP/SVR/GW/height}),
and to the global constraint (\ref{BP/GC/functional/explicit}).

\section{Spherical Shape Equation}
\label{sec/sphericalshapeequation}
\subsection{Variational Principle}
Applying the method of Lagrange multipliers
for differential equation constraints and integral constraints \cite{Pike}
gives the unconstrained functional
\begin{multline}
\label{BP/SVR/Fb/functional/def}
\stGF\left[\stSurfaceOfRevolution\right]%
	=%
	\pi\stKb%
	{\int_{\stProfile}}\!%
	\std{u}\;%
	\left[%
		\bigl(%
			{\partial_{u}\stNPolarAngle(u)}%
		\bigr)^{2}%
		+%
		{\sin^{2}\stNPolarAngle(u)}%
	\right]%
	\\
	+%
	2\pi\stKb%
	{\int_{\stProfile}}\!\!%
	\std{u}\;%
	\stLm(u)\:%
	\bigl[%
		{\cos\stNPolarAngle(u)}%
		-%
		\partial_{u}{\stWeylScalingExp(u)}%
	\bigr]%
	\\
	+%
	2\pi\stKb%
	{\int_{\stProfile}}\!\!%
	\std{u}\;%
	\stLn(u)\:%
	\bigl[%
		{\partial_{u}{z(u)}}%
		+%
		\stE^{\stWeylScalingExp(u)}%
		\sin\stNPolarAngle(u)%
	\bigr]%
	\\
	+%
	2\pi\stKb\:%
	\stLa%
	{\int_{\stProfile}}\!\!%
	\std{u}
	,
	\hphantom{{\partial_{u}{z(u)}}+\stE^{\stWeylScalingExp(u)+\sin\stNPolarAngle(u)}}
\end{multline}
where $\stLm$ and $\stLn$ are local Lagrange multipliers
while $\stLa$ is a global Lagrange multiplier.
Whereas
the Euler-Lagrange equations derived
from the functional (\ref{BP/SVR/Fb/functional/def})
by varying with respect to $\stLm$ and $\stLn$ reproduce, respectively,
the elastic compatibility condition (\ref{BP/SVR/intcond})
and the geometrical confinement (\ref{BP/SVR/GW/height})
as required,
the ones
with respect to $\stWeylScalingExp$, $\stNPolarAngle$, and $z$ are, respectively,
\begin{subequations}
\begin{align}
\label{BP/SVR/Fb/functional/EL/WeylScalingAngle}
	\partial_{u}\stLm(u)&=%
		-%
		\stLn(u)\:%
		\stE^{\stWeylScalingExp(u)}%
		\sin\stNPolarAngle(u)%
	,%
	\\
\label{BP/SVR/Fb/functional/EL/NPolarAngle}
	\begin{split}
	\partial_{uu}\stNPolarAngle(u)&=%
		\sin\stNPolarAngle(u)\cos\stNPolarAngle(u)%
		\\
		&\quad%
		-\stLm(u)\:%
		\sin\stNPolarAngle(u)%
		+\stLn(u)\:%
		\stE^{\stWeylScalingExp(u)}%
		\cos\stNPolarAngle(u)%
	,%
	\end{split}
	\\
\label{BP/SVR/Fb/functional/EL/z}
	\partial_{u}\stLn(u)&=%
		0%
	.
\end{align}
\end{subequations}
Furthermore,
since the integrand of the functional (\ref{BP/SVR/Fb/functional/def})
does not depend explicitly on the variable of integration $u$,
the Beltrami identity \cite{Pike} may be computed;
it yields
\begin{multline}
\label{BP/SVR/Fb/functional/Beltrami}
	-\tfrac{1}{2}%
		\bigl(%
			{\partial_{u}\stNPolarAngle(u)}%
		\bigr)^{2}%
	+%
	\tfrac{1}{2}%
		{\sin^{2}\stNPolarAngle(u)}%
	+%
	\stLm(u)\:%
		{\cos\stNPolarAngle(u)}%
	\\
	+%
	\stLn(u)\:%
		\stE^{\stWeylScalingExp(u)}%
		\sin\stNPolarAngle(u)%
	+%
	\stLa%
	=%
		\stBconst%
	,
\end{multline}
with $\stBconst$ a constant of integration.
On the other hand,
at the boundary ${\partial\stProfile}$,
arbitrary small variations of $\stWeylScalingExp$, $\stNPolarAngle$, $z$, and $u$
---
denoted by
$\delta{\stWeylScalingExp}$, $\delta{\stNPolarAngle}$, $\delta{z}$, and $\delta{u}$,
respectively,
---
must hold:
\begin{subequations}
\begin{align}
\label{BP/SVR/Fb/functional/BC/WeylScalingAngle}
	\Bigl.\stLm(u)\;\delta{\stWeylScalingExp(u)}\Bigr|_{\partial\stProfile}&=0%
	,\\
\label{BP/SVR/Fb/functional/BC/NPolarAngle}
	\Bigl.\partial_{u}{\stNPolarAngle(u)}\;\delta{\stNPolarAngle(u)}\Bigr|_{\partial\stProfile}&=0%
	,\\
\label{BP/SVR/Fb/functional/BC/z}
	\Bigl.\stLn(u)\;\delta{z(u)}\Bigr|_{\partial\stProfile}&=0%
	,\\
\label{BP/SVR/Fb/functional/BC/Beltrami}
	\Bigl.\stBconst\;\delta{u}\Bigr|_{\partial\stProfile}&=0%
	.
\end{align}
\end{subequations}
Clearly the conditions (\ref{BP/SVR/Fb/functional/BC/WeylScalingAngle})
and (\ref{BP/SVR/Fb/functional/BC/NPolarAngle})
are always satisfied at both poles:
at the south pole the radius must vanish ($\delta{\stWeylScalingExp}\!=\!0$)
and the surface normal must stay parallel to the axis of revolution
($\delta{\stNPolarAngle}\!=\!0$),
whereas at the north pole
the grafted latex bead dictates both a fixed radius
($\delta{\stWeylScalingExp}\!=\!0$)
and a fixed polar angle
($\delta{\stNPolarAngle}\!=\!0$)
along the interface parallel \cite{CurveDefinitions}.
Let us now ignore for a while the condition (\ref{BP/SVR/Fb/functional/BC/z})
and focus on the last condition (\ref{BP/SVR/Fb/functional/BC/Beltrami}).
In general,
because there is no restriction on the meridian arc-length between the podal boundaries
\cite{SEASVC}
--- namely because there is no restriction on the isometric coordinate $u$ ---,
the variable $u$ must be permitted to vary freely at the endpoints,
therefore the Beltrami constant of integration $\stBconst$ must be set to zero
in order to fulfill the condition (\ref{BP/SVR/Fb/functional/BC/Beltrami}).

However,
since the imposed global constraint (\ref{BP/GC/functional/def}),
according to Eq.~(\ref{BP/GC/functional/explicit}),
fixes the length of the interval along which the variable $u$ must vary,
the variable $u$ is (intentionally) not allowed to vary freely at the endpoints here.
Thus, here, the Beltrami constant of integration $\stBconst$ is (artificially) an arbitrary constant.
Nonetheless,
we have to remember that the imposed global constraint (\ref{BP/GC/functional/def})
[or (\ref{BP/GC/functional/explicit})]
is meant to mimic true global constraints.
Besides, it is noteworthy
that the Beltrami constant of integration $\stBconst$ can be absorbed
by the artificial global Lagrange multiplier $\stLa$
associated to this global constraint.
Below, with respect to our approach so far,
the Beltrami constant of integration $\stBconst$ will be set to zero
in order to meet the general statement
associated to the condition (\ref{BP/SVR/Fb/functional/BC/Beltrami})
and in such a way that any ``artifact"
introduced by the imposed constraint (\ref{BP/GC/functional/def})
will only be carried out by the global Lagrange multiplier $\stLa$.
Meanwhile,
as an immediate consequence of Eq.~(\ref{BP/SVR/Fb/functional/EL/z}),
the discerning reader has noticed
that the local Lagrange multiplier $\stLn(u)$ is a constant:
\begin{equation}
\label{BP/SVR/Fb/functional/EL/z/int}
	\stLn(u)=\stLnCI
	.
\end{equation}
Promptly,
having still in mind the geometrical confinement (\ref{BP/SVR/GW/height}),
the same reader has furthermore resolved Eq.~(\ref{BP/SVR/Fb/functional/EL/WeylScalingAngle})
--- formally at least:
\begin{equation}
\label{BP/SVR/Fb/functional/EL/WeylScalingAngle/int}
	\stLm(u)=%
		\stLnCI\:%
		z(u)%
		+%
		\stLmCI
	,
\end{equation}
with $\stLmCI$ a constant of integration.
Therefore,
the Beltrami identity (\ref{BP/SVR/Fb/functional/Beltrami}) writes
\begin{multline}
\label{BP/SVR/Fb/functional/Beltrami/generic}
	-\tfrac{1}{2}%
		\bigl(%
			{\partial_{u}\stNPolarAngle(u)}%
		\bigr)^{2}%
	+%
	\tfrac{1}{2}%
		{\sin^{2}\stNPolarAngle(u)}%
	+%
	\stLmCI\:%
		{\cos\stNPolarAngle(u)}%
	\\
	+%
	\stLnCI%
		\bigl[%
			z(u)\,%
			{\cos\stNPolarAngle(u)}%
			+%
			r(u)\,%
			{\sin\stNPolarAngle(u)}%
		\bigr]%
	+%
	\stLa%
	=%
		0%
	,
\end{multline}
where the reparametrization formula (\ref{BP/SVR/GW/radius}) has been used.

\subsection{Generic Shape Equation}
As a matter of fact,
for generic closed surfaces of revolution,
the conditions (\ref{BP/SVR/Fb/functional/BC/WeylScalingAngle})
and (\ref{BP/SVR/Fb/functional/BC/NPolarAngle}) are in general verified
because the local Weyl gauge field $\stWeylScalingExp$
and the polar angle $\stNPolarAngle$
do not vary freely at the endpoints.
As concerns spherical surfaces of revolution,
their arbitrary small variations at each pole have to vanish,
either as required by continuity and symmetry in the absence of any contact,
or as dictated
along the junction interface in the presence of a contact.
Whereas for free toroidal surfaces of revolution,
because the boundaries coincide,
their arbitrary small variations
on each side of the common boundary have to coincide.
By now, the discerning reader has readily checked that
the differentiation of the Beltrami identity 
(\ref{BP/SVR/Fb/functional/Beltrami/generic})
followed by the substitution of the formulas (\ref{BP/SVR/intcond}), 
(\ref{BP/SVR/GW/height}), (\ref{BP/SVR/Fb/functional/EL/z/int}), and 
(\ref{BP/SVR/Fb/functional/EL/WeylScalingAngle/int})
actually leads to Eq.~(\ref{BP/SVR/Fb/functional/EL/NPolarAngle})
as expected.
In other words,
the Beltrami identity (\ref{BP/SVR/Fb/functional/Beltrami/generic})
is the generic shape equation of our system.
As demonstrated in Appendix~\ref{appendix/B},
the generic shape equation (\ref{BP/SVR/Fb/functional/Beltrami/generic})
becomes the customary shape equation found in the literature
when the ``artifact" disappears.
However,
the generic shape equation (\ref{BP/SVR/Fb/functional/Beltrami/generic})
may be specialized to spherical surfaces of revolution as follows.

\begin{figure}[t]
	\begin{center}
		\includegraphics[width=1.0\linewidth]{svdglbes-CurvatureVersusMeridianArcLengthPlots}%
	\end{center}
	\caption{
		The Gaussian curvature $\stGC$
		versus
		the length $\stMeridianArcLength$
		along meridians from the south pole
		for different encapsulated radii $\stBeadEncapsulatedRadius$
		of the northern grafted latex bead:
		the encapsulated radius $\stBeadEncapsulatedRadius$
		increases in the direction of the arrows.
		Dimensionless latex bead radius $\stBeadRadius\!=\!0.15$;
		relative encapsulated radii ${\stBeadEncapsulatedRadius}/{\stBeadRadius}$
		with respect to the arrows:
		$0$,
		$\frac{1}{4}$,
		$\frac{3}{8}$,
		$\frac{1}{2}$,
		$\frac{5}{8}$,
		$\frac{3}{4}$,
		$\frac{7}{8}$,
		and
		$1$.
		The inset shows the geometry at the north pole:
		the dashed circle profiles the latex bead,
		and the bold arc the polar cap imposed by it;
		the arrows indicate the nomenclature:
		the latex bead radius $\stBeadRadius$
		and the encapsulated radius $\stBeadEncapsulatedRadius$.
		}%
\label{fig/CurvatureVersusMeridianArcLengthPlots}
\end{figure}

\subsection{Topological Specialization}
As observed before,
the bending Hamiltonian (\ref{BP/Hb/functional/def})
can be specialized with respect to the underlying topology
through the Bogomol'nyi decomposition technique \cite{BDVAG,SDBTV}:
it appears that the generic shape equation (\ref{BP/SVR/Fb/functional/Beltrami/generic})
can be specialized with respect to the underlying topology as well.
Indeed, the vanishing of the constant of integration $\stLnCI$ depends on
the arbitrariness of the small variations of the height $z$
along the axis of rotation at the boundaries $\partial\stProfile$,
that is to say on the underlying topology.
While for spherical surfaces of revolution
the height $z$ must be free to move along the axis of revolution at each pole,
for free toroidal surfaces of revolution
the small variations of the height $z$ must coincide along the common boundary:
hence in order to meet the boundary condition (\ref{BP/SVR/Fb/functional/BC/z}),
$\stLnCI$ must be zero when the topology is spherical,
whereas $\stLnCI$ is \textit{a priori} arbitrary when the topology is toroidal.
Henceforth,
when the underlying topology is spherical,
the Beltrami identity
or the generic shape equation
(\ref{BP/SVR/Fb/functional/Beltrami/generic}) becomes
\begin{equation}
\label{BP/SVR/Fb/functional/Beltrami/spherical}
	-\tfrac{1}{2}%
		\bigl(%
			{\partial_{u}\stNPolarAngle(u)}%
		\bigr)^{2}%
	+%
	\tfrac{1}{2}%
		{\sin^{2}\stNPolarAngle(u)}%
	+%
	\stLmCI\:%
		{\cos\stNPolarAngle(u)}%
	+%
	\stLa%
	=%
	0
	.
\end{equation}
The reader,
who still has in mind the above implementation of the variational principle,
might point out that this topological specialization allows us
to relax the geometrical confinement (\ref{BP/SVR/GW/height})
as far as only spherical vesicles of revolution
--- governed by an Hamiltonian independent of the height $z$ and its derivatives ---
are concerned.
This approach has been implicitly used by us to derive
the shape equation in the illustration part of a previous Paper
---
namely {Ref.}~\onlinecite{SDBTV}.
Whatever,
the specialized Beltrami identity (\ref{BP/SVR/Fb/functional/Beltrami/spherical})
is the announced spherical shape equation.
Of course,
for more realistic systems the term inherited from global constraints
does not reduce to a constant:
this concern will be the main focus of our future investigations.
As usual,
boundary conditions more specific to the involved system 
allow us to establish relationships between the constants of integration.
Here,
the boundary conditions at the south pole
($\stNPolarAngle\!=\!\pi$, $\partial_{u}{\stNPolarAngle}\!=\!0$)
give
\begin{equation}
	\stLmCI=\stLa%
		\equiv%
		\stDSGConst
	.
\end{equation}
Therefrom,
the spherical Beltrami identity (\ref{BP/SVR/Fb/functional/Beltrami/spherical}) reads
\begin{equation}
\label{BP/SVR/Fb/functional/Beltrami/system}
	-\tfrac{1}{2}%
		\bigl(%
			{\partial_{u}\stNPolarAngle(u)}%
		\bigr)^{2}%
	+%
	\tfrac{1}{2}%
		{\sin^{2}\stNPolarAngle(u)}%
	+%
	\stDSGConst\:%
		{\cos\stNPolarAngle(u)}%
	+%
	\stDSGConst%
	=%
	0
	,
\end{equation}
which is clearly the integrated form of the double sine-Gordon equation
\cite{Leung1982,HSCAMF}
\begin{equation}
\label{BP/SVR/Hb/DSG/equation}
	\partial_{uu}\stNPolarAngle(u)=%
		\sin\stNPolarAngle(u)\,\cos\stNPolarAngle(u)%
		-\stDSGConst\:%
    \sin\stNPolarAngle(u)%
	.
\end{equation}
Let us notice that the coefficient $\stDSGConst$ arises
from both the elastic compatibility (\ref{BP/SVR/intcond})
and the imposed global constraint (\ref{BP/GC/functional/def}),
while it is a constant because
the spherical version of the functional (\ref{BP/SVR/Fb/functional/def})
does not depend explicitly on $\stWeylScalingExp$.
As shown below,
grafting a round latex bead on our bare spherical vesicle of revolution
allows us to impose a nonvanishing coefficient $\stDSGConst$
through the prescribed boundary conditions.
Subsequently,
it might also allow us to acquire both a local and a global
insight into the competition between
bending energy, elastic compatibility, and global constraints
through a model which fortunately is tractable.

\subsection{Naive Analogy}
Physically,
for the surface normal,
the elastic compatibility (\ref{BP/SVR/intcond}) acts
as a uniform external axial field of magnitude $\stKb\stDSGConst$:
indeed for the related effective elastic Hamiltonian,
\begin{equation}
\label{BP/SVR/Heff/canonic/Hamiltonian}
\stEHmBnd\left[\stSurfaceOfRevolution\right]%
	\!=\!%
	\pi\stKb%
	{\int_{\stProfile}}\!%
	\std{u}%
	\left[%
		\bigl(%
			{\partial_{u}\stNPolarAngle(u)}%
		\bigr)^{2}%
		+%
		{\sin^{2}\stNPolarAngle(u)}%
		+%
		2\,%
		\stDSGConst\:%
		{\cos{\stNPolarAngle(u)}}%
	\right]%
	,
\end{equation}
the Euler-Lagrange equation with respect to $\stNPolarAngle$
is the double sine-Gordon equation (\ref{BP/SVR/Hb/DSG/equation});
for spherical surfaces of revolution $\stSurfaceOfRevolution_{0}$,
the effective Hamiltonian (\ref{BP/SVR/Heff/canonic/Hamiltonian})
decomposes
with respect to the Bogomol'nyi technique
into:
\begin{multline}
\label{BP/SVR/Heff/Bogo/decomposition/Hamiltonian}
\stEHmBnd\left[\stSurfaceOfRevolution_{0}\right]%
	=%
	\pi\stKb%
	{\int_{\stProfile}}\!%
	\std{u}\;%
	\bigl[
		\partial_{u}\stNPolarAngle(u)%
		-%
		\sin\stNPolarAngle(u)%
	\bigr]^2%
	\\
	+%
	2\pi\stKb%
	\:\stDSGConst%
	{\int_{\stProfile}}\!%
	\std{u}\;%
		{\cos{\stNPolarAngle(u)}}%
	+%
	4\pi\stKb%
	.
\end{multline}

\section{Local versus Global}
\label{sec/localversusglobal}
\subsection{Exact Solution}
Fortunately
a suitable solution of Eq.~(\ref{BP/SVR/Fb/functional/Beltrami/system}) 
[or (\ref{BP/SVR/Hb/DSG/equation})] 
generalizing the solution (\ref{BP/SVR/Hb/saturation/SG/arctan})
exists
that meets the interface conditions imposed by the grafted latex bead
as follows.
First,
we easily check
by direct substitution
that the solution
\begin{equation}
\label{BP/SVR/Hb/DSG/kink/arctan}
	\stNPolarAngle(u\!\mid\!\stDSGEccentricity)\!=\!%
		2%
		\arctan{%
			\biggl(
				\sqrt{\tfrac{1-\stDSGEccentricity}{1+\stDSGEccentricity}}\;%
				\cosh{\sqrt{\tfrac{1+\stDSGEccentricity}{1-\stDSGEccentricity}}\,u}%
				+%
				\sinh{\sqrt{\tfrac{1+\stDSGEccentricity}{1-\stDSGEccentricity}}\,u}%
			\biggr)%
			}%
\end{equation}
satisfies the second order
nonlinear
differential
equation (\ref{BP/SVR/Hb/DSG/equation})
with
\begin{equation}
\label{BP/SVR/Hb/DSG/kink/constint}
	\stDSGConst=\frac{2\,\stDSGEccentricity}{1-\stDSGEccentricity}%
	\qquad\text{or}\qquad%
	\stDSGEccentricity=\frac{\stDSGConst}{2+\stDSGConst}%
	\:,
\end{equation}
whereas
it reduces to the solution (\ref{BP/SVR/Hb/saturation/SG/arctan})
when the parameter $\stDSGEccentricity$ vanishes.
Solving the relevant Gauss-Weingarten equations
under appropriate conditions (Appendix~\ref{appendix/A}) leads,
after detailed calculations,
to an exact solution for the surface of revolution,
which is the relevant generalization of (\ref{BP/SVR/Hb/saturation/surface}):
\begin{widetext}
\begin{equation}
\label{BP/SVR/Hb/DSG/surface/parametrization}
\begin{split}
	r(u\!\mid\!\stDSGEccentricity)%
	&%
	=%
		4%
		\frac
			{%
				\Bigl(%
					\frac
						{\sqrt{{1+\stDSGEccentricity}}+\sqrt{{1+3\stDSGEccentricity}}}%
						{\sqrt{{1+\stDSGEccentricity}}+\sqrt{{1-\stDSGEccentricity}}}%
				\Bigr)%
				^{\sqrt{\tfrac{1-\stDSGEccentricity}{1+\stDSGEccentricity}}}%
				}
			{%
				\bigl(%
					\scriptstyle{%
						{\sqrt{{1-\stDSGEccentricity}}+\sqrt{{1+3\stDSGEccentricity}}}%
						}
				\bigr)%
				^{2}%
				}%
		\biggl[%
			1%
			+%
			\sqrt{1-\stDSGEccentricity^2}\:%
			\tanh{\sqrt{\tfrac{1+\stDSGEccentricity}{1-\stDSGEccentricity}}\,u}%
		\biggr]%
		\stE^{-u}%
	,%
	\\
	z(u\!\mid\!\stDSGEccentricity)%
	&%
	=%
		\frac{2}{1+\sqrt{\tfrac{1+3\stDSGEccentricity}{1-\stDSGEccentricity}}}%
		\left[%
			2%
			\frac
				{%
					\Bigl(%
						\frac
							{\sqrt{{1+\stDSGEccentricity}}+\sqrt{{1+3\stDSGEccentricity}}}%
							{\sqrt{{1+\stDSGEccentricity}}+\sqrt{{1-\stDSGEccentricity}}}%
					\Bigr)%
					^{\sqrt{\tfrac{1-\stDSGEccentricity}{1+\stDSGEccentricity}}}%
					}
				{%
					\sqrt{\tfrac{1-\stDSGEccentricity}{1+\stDSGEccentricity}}%
					+%
					\sqrt{\tfrac{1+3\stDSGEccentricity}{1+\stDSGEccentricity}}%
					}%
			\:%
			\frac
				{\stE^{-u}}
				{\cosh{\sqrt{\tfrac{1+\stDSGEccentricity}{1-\stDSGEccentricity}}\,u}}%
			-%
			1%
		\right]%
	,
\end{split}
\end{equation}
\end{widetext}
and
which smoothly joins the concavely bound spherical northern (bead) cap
to the vesicle surface when the interface parallel \mbox{$u=\hat{u}$}
\cite{CurveDefinitions}
and the parameter $\stDSGEccentricity$ obey
\begin{equation}
\label{BP/SVR/Hb/DSG/surface/junction}
\begin{gathered}
	0\leqslant\stDSGEccentricity<1%
	,\quad%
	r(\hat{u}\!\mid\!\stDSGEccentricity)=%
		\stBeadEncapsulatedRadius%
	,\quad%
	\stNPolarAngle(\hat{u}\!\mid\!\stDSGEccentricity)=%
		-\arcsin{\frac{\stBeadEncapsulatedRadius}{\stBeadRadius}}%
	,
	\\
	\ln\Bigl(%
		{\tfrac
			{\sqrt{{1+\stDSGEccentricity}}-\sqrt{{1-\stDSGEccentricity}}}%
			{\sqrt{{1+\stDSGEccentricity}}+\sqrt{{1+3\stDSGEccentricity}}}%
			}
	\Bigr)%
	\!\leqslant\!%
	\sqrt{{\tfrac{1+\stDSGEccentricity}{1-\stDSGEccentricity}}}%
	\;%
	\hat{u}%
	\!\leqslant\!%
	{\tfrac{1}{2}}%
	\ln\Bigl(%
		{\tfrac
			{\sqrt{{1+\stDSGEccentricity}}-\sqrt{{1-\stDSGEccentricity}}}%
			{\sqrt{{1+\stDSGEccentricity}}+\sqrt{{1-\stDSGEccentricity}}}%
			}
	\Bigr)%
	,
\end{gathered}
\end{equation}
$\stBeadRadius$ and $\stBeadEncapsulatedRadius$
being, respectively, the latex bead radius and the encapsulated radius,
the isometric coordinate $u$ varying from $\hat{u}$ to $+\infty$
(see {Fig.}~\ref{fig/GlobalCrossSections}
and the inset of {Fig.}~\ref{fig/CurvatureVersusMeridianArcLengthPlots}).
Basic analytical considerations
always
ensure that the system (\ref{BP/SVR/Hb/DSG/surface/junction})
has one unique solution
which can be found numerically without difficulty.
The scale and the position in space
have been chosen
with respect to the equator
\cite{CurveDefinitions}:
its radius is set to unity
(its scale)
while its plane passes through the origin
(its position).
It can be readily verified that
the surface (\ref{BP/SVR/Hb/DSG/surface/parametrization}) 
approaches the round sphere (\ref{BP/SVR/Hb/saturation/surface}),
as expected,
when the parameter $\stDSGEccentricity$ tends to zero.
In fact,
the interface conditions (\ref{BP/SVR/Hb/DSG/surface/junction})
essentially require that at the proper place
(the inequalities)
the normal lines of the vesicle surface and of the round latex 
bead surface coincide
(the equalities)
\cite{Koltover1999,SDBTV,HelfrichKozlov1994}.

When the latex bead is very tightly bound to the vesicle
---
biotinylated lipids can be employed
to render vesicles very sticky
to streptavidin coated latex beads
\cite{Koltover1999}
---
these junction conditions are reasonable at the vesicle scale,
nevertheless at the membrane scale
a more detailed treatment may be needed since
the curvature experiences a discontinuity
along the junction parallel
---
they are the ones implicitly assumed in {Ref.}~\onlinecite{Koltover1999}.
The relative magnitude $\stDSGConst$ of the effective uniform external field,
according to the equality (\ref{BP/SVR/Hb/DSG/kink/constint})
and the interface conditions (\ref{BP/SVR/Hb/DSG/surface/junction}),
is determined by the latex bead radius $\stBeadRadius$
and the encapsulated radius $\stBeadEncapsulatedRadius$:
for physically relevant latex beads,
the relative magnitude $\stDSGConst$ increases as either
the relative encapsulated radius \mbox{${\stBeadEncapsulatedRadius}/{\stBeadRadius}$}
or the latex bead radius $\stBeadRadius$ increases.

\subsection{Hat-Model Limit}
Finally,
for complete encapsulation
(\mbox{${\stBeadEncapsulatedRadius}={\stBeadRadius}$})
in the limit \mbox{$\stDSGEccentricity\ll1$},
the ``hat-model'' approximation
\cite{Koltover1999,HelfrichKozlov1994,Helfrich1998}
is recovered as
the axisymmetric surface (\ref{BP/SVR/Hb/DSG/surface/parametrization})
experiences complete contact of order
\mbox{$\stO\bigl(\stDSGEccentricity^2\ln^2{\stDSGEccentricity}\bigr)$}
with a catenoid along the corresponding interface parallel,
which then fuses itself with the (neck) equator
	\mbox{%
		$%
			u=%
				-%
				\sqrt{\frac{1-\stDSGEccentricity}{1+\stDSGEccentricity}}%
				\ln\Bigl(%
					{\tfrac
						{\sqrt{{1+\stDSGEccentricity}}+\sqrt{{1+3\stDSGEccentricity}}}%
						{\sqrt{{1+\stDSGEccentricity}}-\sqrt{{1-\stDSGEccentricity}}}%
						}
				\Bigr)%
		$%
		}. 
After the coordinate
shift
	\mbox{%
		$%
			u=%
				v%
				-%
				\sqrt{{\tfrac{1-\stDSGEccentricity}{1+\stDSGEccentricity}}}%
				\ln\Bigl(%
					{\tfrac
						{\sqrt{{1+\stDSGEccentricity}}+\sqrt{{1+3\stDSGEccentricity}}}%
						{\sqrt{{1+\stDSGEccentricity}}-\sqrt{{1-\stDSGEccentricity}}}%
						}
				\Bigr)%
		$%
		}%
	,
we get
\begin{align*}
		r(v\!\mid\!\stDSGEccentricity)%
		&%
		=%
			2%
			\stDSGEccentricity\:\cosh{v}%
			+%
			2\,%
			\stDSGEccentricity^2\ln\stDSGEccentricity\:\cosh{v}%
			+\cdots%
		,%
		\\
		z(v\!\mid\!\stDSGEccentricity)%
		&%
		=%
			1%
			+%
			2%
			\stDSGEccentricity\,%
			\bigl(\ln{{\tfrac{1}{2}}\stDSGEccentricity}+{\tfrac{1}{2}}\bigr)%
			+%
			2%
			\stDSGEccentricity\:{v}%
			+%
			\stDSGEccentricity^2\ln^2{\stDSGEccentricity}%
			+\cdots%
		;
\end{align*}
since 
	\mbox{%
		$%
			r(\hat{v}\!=\!0\!\mid\!\stDSGEccentricity)=%
				2\,%
				\stDSGEccentricity%
				+%
				\stO\bigl(%
					\stDSGEccentricity^2\ln^2{\stDSGEccentricity}%
				\bigr)%
		$%
		},
the limit \mbox{$\stDSGEccentricity\ll1$}
is here equivalent to the limit \mbox{$\stBeadRadius\ll1$},
thus it applies to small latex beads.
More detailed analysis may be required for arbitrary encapsulations
(\mbox{${0<\stBeadEncapsulatedRadius}\leqslant{\stBeadRadius}$}).

\subsection{Geometrical Frustration}
Next, while keeping in mind that grafting a second identical latex bead 
antipodal to the first one
(\textit{i.e.}, at the south pole)
cancels the coefficient $\stDSGConst$ in equation (\ref{BP/SVR/Hb/DSG/equation})
as it was implicitly shown in {Ref.}~\onlinecite{SDBTV},
we investigate the global deformation experienced
by our bare spherical vesicle.
First,
for the surface of revolution (\ref{BP/SVR/Hb/DSG/surface/parametrization}),
the length \mbox{$\stMeridianArcLength(u\!\mid\!\stDSGEccentricity)$}
along meridians
\cite{CurveDefinitions}
from the south pole \mbox{$u=+\infty$} to the parallel $u$,
the Gaussian curvature \mbox{$\stGC(u\!\mid\!\stDSGEccentricity)$},
and the mean curvature \mbox{$\stMC(u\!\mid\!\stDSGEccentricity)$}
along the parallel $u$ are found to take, respectively, the exact forms
\begin{widetext}
\begin{align}
\label{BP/SVR/Hb/DSG/surface/arclength/meridian}
	\stMeridianArcLength(u\!\mid\!\stDSGEccentricity)&=%
		2\,%
		\frac
			{%
				\Bigl(%
					\frac
						{\sqrt{{1+\stDSGEccentricity}}+\sqrt{{1+3\stDSGEccentricity}}}%
						{\sqrt{{1+\stDSGEccentricity}}+\sqrt{{1-\stDSGEccentricity}}}%
				\Bigr)%
				^{\sqrt{\tfrac{1-\stDSGEccentricity}{1+\stDSGEccentricity}}}%
				}
			{%
				\Bigl(%
					{1+\sqrt{\tfrac{1+3\stDSGEccentricity}{1-\stDSGEccentricity}}}%
				\Bigr)%
				^{2}%
				}%
		\biggl[%
			\Bigl(%
				{1-\sqrt{\tfrac{1+\stDSGEccentricity}{1-\stDSGEccentricity}}}%
			\Bigr)%
			^{2}%
			+%
			2%
			\LerchPhi\biggl(%
				-\stE^{-2\sqrt{\tfrac{1+\stDSGEccentricity}{1-\stDSGEccentricity}}\,u}%
				,%
				1%
				,%
				\tfrac{1}{2}%
				\sqrt{\tfrac{1-\stDSGEccentricity}{1+\stDSGEccentricity}}%
			\biggr)%
		\biggr]%
		\stE^{-u}%
	,\\
\label{BP/SVR/Hb/DSG/surface/curvature/intrinsic}
	\stGC(u\!\mid\!\stDSGEccentricity)&=%
		\frac{1}{16}%
		\frac
			{%
				(1-\stDSGEccentricity^2)^\frac{3}{2}%
				\Bigl(%
					{1+\sqrt{\tfrac{1+3\stDSGEccentricity}{1-\stDSGEccentricity}}}%
				\Bigr)%
				^{4}%
				}%
			{%
				\Bigl(%
					\frac
						{\sqrt{{1+\stDSGEccentricity}}+\sqrt{{1+3\stDSGEccentricity}}}%
						{\sqrt{{1+\stDSGEccentricity}}+\sqrt{{1-\stDSGEccentricity}}}%
				\Bigr)%
				^{2\sqrt{\tfrac{1-\stDSGEccentricity}{1+\stDSGEccentricity}}}%
				}
		\frac
			{
				\sqrt{1-\stDSGEccentricity^2}%
				\cosh{2\sqrt{\tfrac{1+\stDSGEccentricity}{1-\stDSGEccentricity}}\,u}%
				+%
				\sinh{2\sqrt{\tfrac{1+\stDSGEccentricity}{1-\stDSGEccentricity}}\,u}%
				}
			{
				\biggl[%
					\cosh{\sqrt{\tfrac{1+\stDSGEccentricity}{1-\stDSGEccentricity}}\,u}%
					+%
					\sqrt{{1-\stDSGEccentricity^2}}%
					\sinh{\sqrt{\tfrac{1+\stDSGEccentricity}{1-\stDSGEccentricity}}\,u}%
				\biggr]^4%
				}
		\;%
		\stE^{2u}%
	,\\
\label{BP/SVR/Hb/DSG/surface/curvature/mean}
	\stMC(u\!\mid\!\stDSGEccentricity)&=%
		-%
		\frac{1}{4}%
		\frac
			{%
				(1-\stDSGEccentricity^2)^\frac{1}{2}%
				\Bigl(%
					{1+\sqrt{\tfrac{1+3\stDSGEccentricity}{1-\stDSGEccentricity}}}%
				\Bigr)%
				^{2}%
				}%
			{%
				\Bigl(%
					\frac
						{\sqrt{{1+\stDSGEccentricity}}+\sqrt{{1+3\stDSGEccentricity}}}%
						{\sqrt{{1+\stDSGEccentricity}}+\sqrt{{1-\stDSGEccentricity}}}%
				\Bigr)%
				^{\sqrt{\tfrac{1-\stDSGEccentricity}{1+\stDSGEccentricity}}}%
				}
		\;%
		\frac
			{
				\stE^{u}%
				}
			{
					\cosh{\sqrt{\tfrac{1+\stDSGEccentricity}{1-\stDSGEccentricity}}\,u}%
					+%
					\sqrt{{1-\stDSGEccentricity^2}}%
					\sinh{\sqrt{\tfrac{1+\stDSGEccentricity}{1-\stDSGEccentricity}}\,u}%
				}
	,
\end{align}
\end{widetext}
with $\LerchPhi$ denoting the Lerch transcendental function
\cite{TISP}.
Plotting the Gaussian curvature $\stGC$
as a function of the meridian arclength $\stMeridianArcLength$
from the south pole reveals, as clearly exhibited in 
{Fig.}~\ref{fig/CurvatureVersusMeridianArcLengthPlots},
that the surface of our bare vesicle distorted by a grafted round latex bead
surprisingly splits into three domains:
a strong curvature gradient region in the neighborhood of the bead;
a weak curvature gradient zone around the equator;
and an unexpectedly flat southern region,
as the curvature amazingly tends to zero at the south pole.
On the other hand,
the presence (or absence) of an antipodal grafted latex bead
does not seem to radically affect the profiles around the grafted latex bead,
as is evident by comparing {Fig.}~\ref{fig/GlobalCrossSections}
with the corresponding one in {Ref.}~\onlinecite{SDBTV}.
The global deformation of our bare spherical vesicle
can be understood as follows.
By imposing a concavely bound spherical cap,
the grafted round latex bead
prevents the bending Hamiltonian (\ref{BP/SVR/Hb/canonic/Hamiltonian})
[or (\ref{BP/SVR/Hb/Bogo/decomposition/Hamiltonian})] 
from reaching its global minimum
and brings into play the elastic compatibility (\ref{BP/SVR/intcond}):
the former phenomenon is revealed through the Bogomol'nyi decomposition technique,
the latter can be mimicked by an effective uniform external axial field
interacting with the surface normal.

This key result concisely expressed
in the decomposed effective Hamiltonian
(\ref{BP/SVR/Heff/Bogo/decomposition/Hamiltonian})
leads to a clear comprehension of
the geometrical frustration experienced by our bare spherical vesicle:
the surface normal attempts
to shape our bare spherical vesicle into a round sphere
to saturate that part of the effective Hamiltonian
(\ref{BP/SVR/Heff/Bogo/decomposition/Hamiltonian})
which is inherited from the bending Hamiltonian,
whereas
the effective uniform external axial field
tries to antialign the surface normal
to minimize that part of the effective Hamiltonian
(\ref{BP/SVR/Heff/Bogo/decomposition/Hamiltonian})
which mimics a uniform external axial interaction.
Ultimately the alteration due to the presence (or absence)
of an antipodal grafted latex bead
resolves the competition between the two parts:
the highly distorted region in the neighborhood of 
the grafted bead is mainly governed by the shape part
which tends to make it round;
by contrast,
the essentially flat region antipodal to the grafted bead
is driven by the effective uniform external axial field
which flattens it by forcing the local curvature to vanish 
while the shape part tries to maintain it round. 
The elastic compatibility condition (\ref{BP/SVR/intcond})
globally ``releases''
the bending Hamiltonian (\ref{BP/SVR/Hb/canonic/Hamiltonian})
[or (\ref{BP/SVR/Hb/Bogo/decomposition/Hamiltonian})] 
in the sense that omitting the elastic compatibility condition 
(\ref{BP/SVR/intcond}) is tantamount to enhancing the 
bending energy by attaching an antipodal latex bead.

\section{Conclusion}
\label{sec/conclusion}
In conclusion,
we have demonstrated that bare spherical vesicles 
distorted by a grafted latex bead might be treated
by taking into account nonlinearity
in the bending elasticity through
the Bogomol'nyi decomposition technique
and elastic compatibility.
We have obtained and analyzed a suitable exact solution,
which recovers the hat-model approximation
\cite{Koltover1999,HelfrichKozlov1994}
in the limit of small latex beads,
through an effective field theory.
The present study provides
a physically motivated example of pure geometrical frustration
induced by a mesoscopic particle.
Finally,
whereas mesoscopic inclusion problems are typically treated
by considering spherical biological vesicles as infinite planes
\cite{Goulian1993,Koltover1999,HelfrichKozlov1994,Helfrich1998,Deserno2004},
we have shown that simple topological arguments may allow us to tackle 
such problems by emphasizing the underlying spherical topology.
We have also demonstrated how our approach can recover the relevant shape equation.
Nevertheless,
in order to overcome one drawback of our exact solution
---
\textit{viz.}, the use of an unphysical global constraint
which is meant to mimic the usual physical global constraints
\mbox{---,}
our suggested approach may be further augmented by accounting for 
the physical global constraints
---
namely, the total mean curvature, 
area, and volume constraints \cite{CFMV,STVPDSCBCM}
---
and, eventually, 
by connecting it to other geometrical approaches
\cite{Gozdz2007,Guven2004,Capovilla2005,CastroVillarrealGuven200703,CastroVillarrealGuven200707}.
Also, by applying it to vesicle systems exhibiting pertinent global 
distortions
---
such as toroidal vesicles \cite{Zhongcan},
vesicle systems exhibiting domains \cite{Baumgart2003},
vesicles distorted by adsorbed objects \cite{Koltover1999},
cylindrical and other inclusions in vesicles 
\cite{Rosso2003,BiscariBisi2002,BiscariNapoli2005,Weikl2003,ZhangMa200606,ZhangMa200612},
and so forth
---,
our suggested approach might lead to more valuable solutions,
but they are difficult to find analytically.

Partial support to J. B. from EC under Contract No. HPRN-CR-1999-00163 (LOCNET network)
and from the National Center for Theoretical Sciences, Taiwan, are acknowledged.
The work at Los Alamos was supported by the US Department of Energy.

\appendix

\section{Surfaces of Revolution in Isometric Azimuthal Coordinates}
\label{appendix/A}
\subsection{Metric and Shape Tensors}
Let us represent any surface of revolution in azimuthal coordinates $(v,\varphi)$ by
\begin{equation}
	\vect{X}(v,\varphi)=%
		\bigl[%
			r(v)\cos\varphi,
			r(v)\sin\varphi,
			z(v)
		\bigr]%
\end{equation}
where the radius $r(v)$ $[{0}\leqslant{r(v)}]$
and
the height $z(v)$
are both sufficiently differentiable functions of the coordinate $v$.
Then the metric tensor ${\tens{g}}_{ij}$ has the following components
\cite{Struik,GPI}:
\begin{subequations}
\begin{align}
	\begin{split}
	{\tens{g}}_{vv}&=%
		\partial_{v}\vect{X}(v,\varphi)\cdot\partial_{v}\vect{X}(v,\varphi)%
		\\
		&=%
		\bigl(\partial_{v}{r(v)}\bigr)^{2}%
		+%
		\bigl(\partial_{v}{z(v)}\bigr)^{2}
		,%
	\end{split}
	\\
	{\tens{g}}_{v\varphi}&={\tens{g}}_{{\varphi}v}=%
		\partial_{v}\vect{X}(v,\varphi)\cdot\partial_{\varphi}\vect{X}(v,\varphi)=%
		0%
	,%
	\\
	{\tens{g}}_{\varphi\varphi}&=%
		\partial_{\varphi}\vect{X}(v,\varphi)\cdot\partial_{\varphi}\vect{X}(v,\varphi)=%
		r(v)^{2}%
	.
\end{align}
\end{subequations}
Before computing the shape tensor ${\tens{b}}_{ij}$,
the outward surface unit normal vector $\widehat{\vect{N}}(v,\varphi)$ may be computed as
\cite{Struik,GPI}
\begin{equation}
\begin{split}
	&\widehat{\vect{N}}(v,\varphi)%
		=%
			\frac{%
				\partial_{v}\vect{X}(v,\varphi)\times\partial_{\varphi}\vect{X}(v,\varphi)%
				}{%
				\bigl\Vert%
					\partial_{v}\vect{X}(v,\varphi)\times\partial_{\varphi}\vect{X}(v,\varphi)%
				\bigr\Vert%
				}
		\\
		&\quad%
		=%
				\frac{
					1%
					}{%
					\sqrt{%
						\bigl(\partial_{v}{r(v)}\bigr)^{2}%
						+%
						\bigl(\partial_{v}{z(v)}\bigr)^{2}%
						}
					}%
				\\
				&\qquad\qquad\cdot%
				\bigl[%
					-\partial_{v}{z(v)}\cos\varphi,
					-\partial_{v}{z(v)}\sin\varphi,
					\partial_{v}{r(v)}
				\bigr]%
		.
\end{split}
\end{equation}
Hence \cite{Struik,GPI}
\begin{subequations}
\begin{align}
	\begin{split}
	{\tens{b}}_{vv}&=%
		\partial_{vv}\vect{X}(v,\varphi)\cdot\widehat{\vect{N}}(v,\varphi)%
		\\
		&=%
		\frac{
			\partial_{vv}{z(v)}\partial_{v}{r(v)}%
			-%
			\partial_{vv}{r(v)}\partial_{v}{z(v)}%
			}{%
			\sqrt{%
				\bigl(\partial_{v}{r(v)}\bigr)^{2}%
				+%
				\bigl(\partial_{v}{z(v)}\bigr)^{2}%
				}
			}
	,%
	\end{split}
	\\
	{\tens{b}}_{v\varphi}&={\tens{b}}_{{\varphi}v}=%
		\partial_{v\varphi}\vect{X}(v,\varphi)\cdot\widehat{\vect{N}}(v,\varphi)=%
		0
	,%
	\\
	\begin{split}
	{\tens{b}}_{\varphi\varphi}&=%
		\partial_{\varphi\varphi}\vect{X}(v,\varphi)\cdot\widehat{\vect{N}}(v,\varphi)%
		\\
		&=%
		\frac{
			{r(v)}\partial_{v}{z(v)}%
			}{%
			\sqrt{%
				\bigl(\partial_{v}{r(v)}\bigr)^{2}%
				+%
				\bigl(\partial_{v}{z(v)}\bigr)^{2}%
				}
			}
		.
	\end{split}
\end{align}
\end{subequations}
Next we may introduce the polar angle $\stNPolarAngle(v)$
associated with $\widehat{\vect{N}}(v,\varphi)$:
\begin{equation}
	\stNPolarAngle(v)=%
		\arctan\bigl(%
			-\partial_{v}{z(v)}
			,
			\partial_{v}{r(v)}
			\bigr)%
	,
\end{equation}
which satisfies the (useful) relationships:
\begin{subequations}
\begin{align}
	\cos\stNPolarAngle(v)&=%
		\frac{
			\partial_{v}{r(v)}%
			}{%
			\sqrt{%
				\bigl(\partial_{v}{r(v)}\bigr)^{2}%
				+%
				\bigl(\partial_{v}{z(v)}\bigr)^{2}%
				}%
			}%
	,
	\\
	\sin\stNPolarAngle(v)&=%
		\frac{
			-\partial_{v}{z(v)}%
			}{%
			\sqrt{%
				\bigl(\partial_{v}{r(v)}\bigr)^{2}%
				+%
				\bigl(\partial_{v}{z(v)}\bigr)^{2}%
				}%
			}%
	,%
	\\
	\partial_{v}\stNPolarAngle(v)&=%
		\frac{
			\partial_{vv}{r(v)}\partial_{v}{z(v)}%
			-%
			\partial_{vv}{z(v)}\partial_{v}{r(v)}%
			}{%
			{%
				\bigl(\partial_{v}{r(v)}\bigr)^{2}%
				+%
				\bigl(\partial_{v}{z(v)}\bigr)^{2}%
				}%
			}%
	.
\end{align}
\end{subequations}
Thus,
whereas the outward surface unit normal vector $\widehat{\vect{N}}(v,\varphi)$
takes the desired form
\begin{equation}
	\widehat{\vect{N}}(v,\varphi)=%
		\bigl[%
			\sin\stNPolarAngle(v)\cos\varphi,
			\sin\stNPolarAngle(v)\sin\varphi,
			\cos\stNPolarAngle(v)
		\bigr]%
	,
\end{equation}
the components of the shape tensor ${\tens{b}}_{ij}$ are readily written as
\begin{subequations}
\begin{align}
	{\tens{b}}_{vv}&=%
		-%
		\sqrt{%
			\bigl(\partial_{v}{r(v)}\bigr)^{2}%
			+%
			\bigl(\partial_{v}{z(v)}\bigr)^{2}
			}%
		\:%
		\partial_{v}\stNPolarAngle(v)%
	,
	\\
	{\tens{b}}_{v\varphi}&={\tens{b}}_{{\varphi}v}=%
		0
	,
	\\
	{\tens{b}}_{\varphi\varphi}&=%
		-%
		r(v)%
		\sin\stNPolarAngle(v)%
	.
\end{align}
\end{subequations}
Now let us transform to isometric coordinates $(u,\varphi)$
\cite{Struik,Nakahara}
by the transformation
\begin{equation}
	u=\int^{v}\!\!\!\!%
		\std{w}%
		\;%
		{\frac{%
			\sqrt{%
				(\partial_{w}{r(w)})^{2}%
				+%
				(\partial_{w}{z(w)})^{2}%
				}
			}{%
			r(w)
			}}
		,
\end{equation}
which yields a conformally flat metric
\cite{Nakahara}.
As a matter of fact,
the components of the metric tensor become
\begin{subequations}
\begin{align}
	{\tens{g}}_{uu}&=%
		r(u)^{2}%
	,
	\\
	{\tens{g}}_{u\varphi}&={\tens{g}}_{{\varphi}{u}}=%
		0
	,
	\\
	{\tens{g}}_{\varphi\varphi}&=%
		r(u)^{2}%
	,
\end{align}
\end{subequations}
while the components of the shape tensor attain the form
\begin{subequations}
\begin{align}
	{\tens{b}}_{uu}&=%
		-%
		r(u)%
		\,%
		\partial_{u}\stNPolarAngle(u)%
	,
	\\
	{\tens{b}}_{u\varphi}&={\tens{b}}_{{\varphi}{u}}=%
		0
	,
	\\
	{\tens{b}}_{\varphi\varphi}&=%
		-%
		r(u)%
		\sin\stNPolarAngle(u)%
	.
\end{align}
\end{subequations}
Finally,
to emphasize the conformally flat nature of the metric
and/or the isometric nature of the coordinates $(u,\varphi)$,
we may introduce the local Weyl gauge field $\stWeylScalingExp(u)$
\cite{Nakahara}:
\begin{equation}
	\stWeylScalingExp(u)=%
		\ln(r(u))
	.
\end{equation}
To summarize,
for any surface of revolution
the metric tensor ${\tens{g}}_{ij}$
and the shape tensor ${\tens{b}}_{ij}$
can take,
in (isometric) azimuthal coordinates $(u,\varphi)$,
the form
\begin{gather}
\label{appendix/tensor/metric/components}
	{\tens{g}}_{uu}={\tens{g}}_{\varphi\varphi}=%
		\stE^{2\stWeylScalingExp(u)}%
	,
	\\
\label{appendix/tensor/shape/components}
	{\tens{b}}_{uu}=%
		-%
		\stE^{\stWeylScalingExp(u)}%
		\partial_{u}\stNPolarAngle(u)%
	\;\text{and}\;%
	{\tens{b}}_{\varphi\varphi}=%
		-%
		\stE^{\stWeylScalingExp(u)}%
		\sin\stNPolarAngle(u)%
	,
\end{gather}
respectively.

\subsection{Elastic Compatibility Conditions}
Next, with the perspective of computing the elastic compatibility conditions
(the Gauss-Codazzi-Peterson equations)
\cite{GPI}
we may compute the Christoffel symbols of the second kind $\Gamma^{k}_{\hphantom{k}{ij}}$
\cite{GPI,Nakahara},
\begin{equation}
	\Gamma^{k}_{\hphantom{x}{ij}}=%
		\tfrac{1}{2}%
		\:%
		{\tens{g}}^{kl}%
			\Bigl[%
				\partial_{i}{\tens{g}}_{lj}%
				+%
				\partial_{j}{\tens{g}}_{li}%
				-%
				\partial_{l}{\tens{g}}_{ij}%
			\Bigr]%
	,
\end{equation}
and then the Riemann tensor ${\tens{R}}_{ijkl}$
\cite{GPI,Nakahara},
\begin{equation}
	{\tens{R}}_{ijkl}=%
		{\tens{g}}_{im}%
			\Bigl[%
				\partial_{k}\Gamma^{m}_{\hphantom{m}{lj}}%
				-%
				\partial_{l}\Gamma^{m}_{\hphantom{m}{kj}}%
				+%
				\Gamma^{m}_{\hphantom{k}{kn}}\Gamma^{n}_{\hphantom{n}{lj}}%
				-%
				\Gamma^{m}_{\hphantom{k}{ln}}\Gamma^{n}_{\hphantom{n}{kj}}%
			\Bigr]%
	.
\end{equation}
In isometric azimuthal coordinates $(u,\varphi)$,
the non-vanishing Christoffel symbols
of the second kind
$\Gamma^{k}_{\hphantom{k}{ij}}$ verify
\begin{equation}
	\Gamma^{u}_{\hphantom{x}{uu}}(u)=%
	-\Gamma^{u}_{\hphantom{x}{{\varphi}{\varphi}}}(u)=%
	\Gamma^{\varphi}_{\hphantom{x}{{u}{\varphi}}}(u)=%
	\Gamma^{\varphi}_{\hphantom{x}{{\varphi}{u}}}(u)=%
	\partial_{u}\stWeylScalingExp(u)%
	,
\end{equation}
and the nonvanishing component of the Riemann tensor ${\tens{R}}_{ijkl}$ is
\begin{equation}
\label{appendix/tensor/riemann}
	{\tens{R}}_{{u}{\varphi}{u}{\varphi}}=%
		-%
		\stE^{2\stWeylScalingExp(u)}%
		\:%
		\partial_{uu}\stWeylScalingExp(u)%
	.
\end{equation}
Therefore
the Gauss equations \cite{GPI}
\begin{equation}
\label{appendix/intcond/gauss/generic}
	{\tens{R}}_{ijkl}=%
		{\tens{b}}_{ik}{\tens{b}}_{jl}%
		-%
		{\tens{b}}_{il}{\tens{b}}_{jk}%
	,
\end{equation}
simplify to the equation
\begin{equation}
\label{appendix/intcond/gauss}
	\partial_{uu}\stWeylScalingExp(u)=%
	-%
	\sin\stNPolarAngle(u)%
	\:%
	\partial_{u}\stNPolarAngle(u)%
	,
\end{equation}
while the equations of Codazzi and Peterson \cite{GPI},
\begin{equation}
\label{appendix/intcond/codazzi/generic}
	\partial_{k}{\tens{b}}_{ij}%
	-%
	\Gamma^{m}_{\hphantom{m}{ik}}{\tens{b}}_{mj}%
	=%
	\partial_{j}{\tens{b}}_{ik}%
	-%
	\Gamma^{n}_{\hphantom{n}{ij}}{\tens{b}}_{nk}%
	,
\end{equation}
yield
\begin{equation}
\label{appendix/intcond/codazzi}
	\partial_{u}\stWeylScalingExp(u)=%
		\cos\stNPolarAngle(u)%
	,
\end{equation}
which clearly satisfies the Gauss equation (\ref{appendix/intcond/gauss}).
In short,
the elastic compatibility conditions
(\ref{appendix/intcond/gauss/generic})--(\ref{appendix/intcond/codazzi/generic})
associated with 
the metric tensor (\ref{appendix/tensor/metric/components})
coupled to the shape tensor (\ref{appendix/tensor/shape/components})
reduce to Eq.~(\ref{appendix/intcond/codazzi}).

\subsection{Fundamental Theorem of Surface Theory}
Next
we shall show the converse of the previous results,
namely,
that any pair of diagonal second-rank tensors \mbox{$({\tens{g}}_{ij},{\tens{b}}_{ij})$}
which in (isometric) azimuthal coordinates \mbox{$(u,\varphi)$}
takes the form (\ref{appendix/tensor/metric/components})--(\ref{appendix/tensor/shape/components})
and obeys the elastic compatibility condition (\ref{appendix/intcond/codazzi}),
where the local Weyl gauge field $\stWeylScalingExp$
and the polar angle $\stNPolarAngle$ of the outward surface normal
are sufficiently  differentiable functions of $u$,
corresponds to a unique axisymmetric surface 
---
modulo its position in space.
In fact, this converse theorem is rather
a simple illustration of the fundamental theorem of surface theory
\cite{Struik,GPI}.
The general demonstration of the fundamental theorem of surface theory
\cite{Struik}
consists in proving that the Gauss surface equations \cite{Struik,GPI}
\begin{equation}
\label{appendix/converse/gauss/generic}
	\partial_{ij}\vect{X}=%
		\Gamma^{m}_{\hphantom{m}{ij}}\:%
			\partial_{m}\vect{X}%
		+%
		{\tens{b}}_{ij}\widehat{\vect{N}}%
	,
\end{equation}
combined with the Weingarten equations \cite{Struik,GPI}
\begin{equation}
\label{appendix/converse/weingarten/generic}
	\partial_{i}\widehat{\vect{N}}=%
		-{\tens{b}}^{m}_{\hphantom{m}{i}}\:%
			\partial_{m}\vect{X}%
	,
\end{equation}
under the additional conditions \cite{Struik}
\begin{equation}
\begin{aligned}
\label{appendix/converse/addcond/generic}
	{\widehat{\vect{N}}}^{2}&=1%
	,%
	&%
	\partial_{i}\vect{X}&\cdot\widehat{\vect{N}}=0%
	,%
	\\
	\partial_{i}\vect{X}&\cdot\partial_{j}\vect{X}={\tens{g}}_{ij}%
	,%
	&%
	\partial_{ij}\vect{X}&\cdot\widehat{\vect{N}}={\tens{b}}_{ij}%
	,
\end{aligned}
\end{equation}
determine a unique surface $\vect{X}$
--- modulo its position in space ---
with ${\widehat{\vect{N}}}$ as its outward surface unit normal vector.
In our case,
the Gauss surface equations (\ref{appendix/converse/gauss/generic}) reduce to
\begin{subequations}
\label{appendix/converse/sode/first}
\begin{align}
\label{appendix/converse/gauss/uu}
	\begin{split}
	\partial_{uu}{\vect{X}(u,\varphi)}&=%
		+\cos\stNPolarAngle(u)%
			\:%
			\partial_{u}\vect{X}(u,\varphi)%
		\\
		&\qquad\qquad%
		-%
		\stE^{\stWeylScalingExp(u)}%
		\partial_{u}\stNPolarAngle(u)%
			\:%
			{\widehat{\vect{N}}(u,\varphi)}
	,%
	\end{split}
	\\
\label{appendix/converse/gauss/uphi}
	\partial_{u\varphi}{\vect{X}(u,\varphi)}&=%
		+\cos\stNPolarAngle(u)%
			\:%
			\partial_{\varphi}\vect{X}(u,\varphi)
	,
	\\
\label{appendix/converse/gauss/phiphi}
	\begin{split}
	\partial_{\varphi\varphi}{\vect{X}(u,\varphi)}&=%
		-\cos\stNPolarAngle(u)%
			\:%
			\partial_{u}\vect{X}(u,\varphi)%
		\\
		&\qquad\qquad%
		-%
		\stE^{\stWeylScalingExp(u)}%
		\sin\stNPolarAngle(u)%
			\:%
			{\widehat{\vect{N}}(u,\varphi)}
	,
	\end{split}
\end{align}
\end{subequations}
and the Weingarten equations (\ref{appendix/converse/weingarten/generic}) to
\begin{subequations}
\begin{align}
\label{appendix/converse/weingarten/u}
	\partial_{u}{\widehat{\vect{N}}(u,\varphi)}&=%
		\stE^{-\stWeylScalingExp(u)}%
		\partial_{u}\stNPolarAngle(u)%
			\:%
			\partial_{u}{\vect{X}(u,\varphi)}
	,
	\\
\label{appendix/converse/weingarten/phi}
	\partial_{\varphi}{\widehat{\vect{N}}(u,\varphi)}&=%
		\stE^{-\stWeylScalingExp(u)}%
		\sin\stNPolarAngle(u)%
			\:%
			\partial_{\varphi}{\vect{X}(u,\varphi)}
	,
\end{align}
\end{subequations}
while the additional conditions (\ref{appendix/converse/addcond/generic}) yield
\begin{subequations}
\label{appendix/converse/addcond/specific}
\begin{align}
\label{appendix/converse/addcond/N}
	{\widehat{\vect{N}}}^{2}(u,\varphi)&=1%
	,
	\\
\label{appendix/converse/addcond/Xu/N}
	\partial_{u}\vect{X}(u,\varphi)\cdot\widehat{\vect{N}}(u,\varphi)&=%
		0%
	,
	\\
\label{appendix/converse/addcond/Xphi/N}
	\partial_{\varphi}\vect{X}(u,\varphi)\cdot\widehat{\vect{N}}(u,\varphi)&=%
		0%
	,
	\\
\label{appendix/converse/addcond/Xu/Xu}
	\partial_{u}\vect{X}(u,\varphi)\cdot\partial_{u}\vect{X}(u,\varphi)&=%
		\stE^{2\stWeylScalingExp(u)}%
	,
	\\
\label{appendix/converse/addcond/Xu/Xphi}
	\partial_{u}\vect{X}(u,\varphi)\cdot\partial_{\varphi}\vect{X}(u,\varphi)&=%
		0%
	,
	\\
\label{appendix/converse/addcond/Xphi/Xphi}
	\partial_{\varphi}\vect{X}(u,\varphi)\cdot\partial_{\varphi}\vect{X}(u,\varphi)&=%
		\stE^{2\stWeylScalingExp(u)}%
	,
	\\
\label{appendix/converse/addcond/Xuu/N}
	\partial_{uu}\vect{X}(u,\varphi)\cdot\widehat{\vect{N}}(u,\varphi)&=%
		-%
		\stE^{\stWeylScalingExp(u)}%
		\partial_{u}\stNPolarAngle(u)%
	,
	\\
\label{appendix/converse/addcond/Xuphi/N}
	\partial_{{u}{\varphi}}\vect{X}(u,\varphi)\cdot\widehat{\vect{N}}(u,\varphi)&=%
		0%
	,
	\\
\label{appendix/converse/addcond/Xphiphi/N}
	\partial_{\varphi\varphi}\vect{X}(u,\varphi)\cdot\widehat{\vect{N}}(u,\varphi)&=%
		-%
		\stE^{\stWeylScalingExp(u)}%
		\sin\stNPolarAngle(u)%
	.
\end{align}
\label{appendix/converse/sode/last}
\end{subequations}
Note that,
provided that conditions
(\ref{appendix/converse/addcond/N})--(\ref{appendix/converse/addcond/Xphi/N}) are met,
conditions
(\ref{appendix/converse/addcond/Xuu/N})--(\ref{appendix/converse/addcond/Xphiphi/N})
are automatically satisfied 
since then they straightaway follow
from Eqs.~(\ref{appendix/converse/gauss/uu})--(\ref{appendix/converse/gauss/phiphi}),
respectively.
By definition,
on the other hand,
the outward surface unit normal vector $\widehat{\vect{N}}(u,\varphi)$ expresses as
\begin{equation}
	\widehat{\vect{N}}(u,\varphi)%
		=%
			\frac{%
				\partial_{u}\vect{X}(u,\varphi)\times\partial_{\varphi}\vect{X}(u,\varphi)%
				}{%
				\bigl\Vert%
					\partial_{u}\vect{X}(u,\varphi)\times\partial_{\varphi}\vect{X}(u,\varphi)%
				\bigr\Vert%
				}%
	,
\end{equation}
which, with respect to conditions
(\ref{appendix/converse/addcond/Xu/Xu})--(\ref{appendix/converse/addcond/Xphi/Xphi}),
writes
\begin{equation}
	\widehat{\vect{N}}(u,\varphi)%
		=%
		\stE^{-2\stWeylScalingExp(u)}\;%
		\partial_{u}\vect{X}(u,\varphi)\times\partial_{\varphi}\vect{X}(u,\varphi)%
	,
\end{equation}
as we find
\begin{equation}
	\bigl[%
		\partial_{u}\vect{X}(u,\varphi)\times\partial_{\varphi}\vect{X}(u,\varphi)%
	\big]^{2}%
		=%
		\stE^{4\stWeylScalingExp(u)}%
	.
\end{equation}
Conditions (\ref{appendix/converse/addcond/N})--(\ref{appendix/converse/addcond/Xphi/N})
are clearly fulfilled.
In short,
only the additional conditions 
(\ref{appendix/converse/addcond/Xu/Xu})--(\ref{appendix/converse/addcond/Xphi/Xphi})
may be pertinent in due course.
By elimination of ${\widehat{\vect{N}}(u,\varphi)}$,
Eqs. (\ref{appendix/converse/gauss/phiphi}), (\ref{appendix/converse/gauss/uphi}),
and (\ref{appendix/converse/weingarten/phi})
hold:
\begin{equation}
\label{appendix/reconstruction/X/step/first/preliminary}
	\partial_{\varphi\varphi\varphi}{\vect{X}(u,\varphi)}%
	+%
	\partial_{\varphi}{\vect{X}(u,\varphi)}%
	=0
	,
\end{equation}
or
\begin{equation}
\label{appendix/reconstruction/X/step/first}
	{\vect{X}(u,\varphi)}=%
		\cos\varphi%
			\:%
			{\vect{A}(u)}%
		+%
		\sin\varphi%
			\:%
			{\vect{B}(u)}%
		+%
			{\vect{C}(u)}
	.
\end{equation}
Then equation (\ref{appendix/converse/gauss/uphi}) gives
\begin{equation}
	\partial_{u}{\vect{A}(u)}=%
		\partial_{u}\stWeylScalingExp(u)%
		\:%
		{\vect{A}(u)}%
	\;\text{and}\;%
	\partial_{u}{\vect{B}(u)}=%
		\partial_{u}\stWeylScalingExp(u)%
		\:%
		{\vect{B}(u)}
		,
\end{equation}
or
\begin{equation}
	{\vect{A}(u)}=%
		\stE^{\stWeylScalingExp(u)}%
		\:%
		{\vect{A}}%
	\quad\text{and}\quad%
	{\vect{B}(u)}=%
		\stE^{\stWeylScalingExp(u)}%
		\:%
		{\vect{B}}
	,
\end{equation}
where ${\vect{A}}$ and ${\vect{B}}$ are constant vectors of integration.
Also, formula (\ref{appendix/reconstruction/X/step/first}) writes
\begin{equation}
\label{appendix/reconstruction/X/step/second}
	{\vect{X}(u,\varphi)}=%
		\stE^{\stWeylScalingExp(u)}%
		\bigl[%
			\cos\varphi%
				\:%
				{\vect{A}}%
			+%
			\sin\varphi%
				\:%
				{\vect{B}}%
		\bigr]%
		+%
		{\vect{C}(u)}
	.
\end{equation}
Moreover,
eliminating ${\widehat{\vect{N}}(u,\varphi)}$
between Eqs. (\ref{appendix/converse/gauss/uu}) and (\ref{appendix/converse/gauss/phiphi}),
we obtain
\begin{multline}
\label{appendix/reconstruction/X/step/third/preliminary}
	\sin\stNPolarAngle(u)%
		\:%
		\partial_{uu}{\vect{X}(u,\varphi)}%
	-%
	\partial_{u}\stNPolarAngle(u)%
		\:%
		\partial_{\varphi\varphi}{\vect{X}(u,\varphi)}%
	\\
	=%
	\cos\stNPolarAngle(u)%
	\,%
	\bigl(%
		\sin\stNPolarAngle(u)%
		+%
		\partial_{u}\stNPolarAngle(u)%
	\bigr)%
		\:%
		\partial_{u}{\vect{X}(u,\varphi)}
	.
\end{multline}
Substitution of formula (\ref{appendix/reconstruction/X/step/second})
into Eq.~(\ref{appendix/reconstruction/X/step/third/preliminary})
yields
\begin{equation}
	\partial_{uu}{\vect{C}(u)}=%
		\bigl[%
			\partial_{u}\stWeylScalingExp(u)%
			+%
			\cot\stNPolarAngle(u)%
			\,%
			\partial_{u}\stNPolarAngle(u)%
		\bigr]%
		\:%
		\partial_{u}{\vect{C}(u)}
	,
\end{equation}
or
\begin{equation}
\label{appendix/reconstruction/X/step/third/amble}
	\partial_{u}{\vect{C}(u)}=%
		-%
		\stE^{\stWeylScalingExp(u)}%
		\sin\stNPolarAngle(u)%
		\:%
		{\vect{C}}
	,
\end{equation}
where ${\vect{C}}$ is a constant vector of integration.
So, we readily have
\begin{equation}
	{\vect{C}(u)}=%
		-%
		\int^{u}\!\!\!\!%
			\std{w}%
			\;%
			\stE^{\stWeylScalingExp(w)}%
			\sin\stNPolarAngle(w)%
			\:%
			{\vect{C}}%
		+%
			{\vect{D}}
	,
\end{equation}
with ${\vect{D}}$ a constant vector of integration.
Henceforth,
formula (\ref{appendix/reconstruction/X/step/second}) becomes
\begin{multline}
\label{appendix/reconstruction/X/step/third}
	{\vect{X}(u,\varphi)}=%
		\stE^{\stWeylScalingExp(u)}%
		\bigl[%
			\cos\varphi%
				\:%
				{\vect{A}}%
			+%
			\sin\varphi%
				\:%
				{\vect{B}}%
		\bigr]%
		\\
		-%
		\int^{u}\!\!\!\!%
			\std{w}%
			\;%
			\stE^{\stWeylScalingExp(w)}%
			\sin\stNPolarAngle(w)%
			\:%
			{\vect{C}}%
		+%
			{\vect{D}}
	.
\end{multline}
We must next choose the constant vectors of integration
${\vect{A}}$, ${\vect{B}}$, ${\vect{C}}$, and ${\vect{D}}$
so as to satisfy the pertinent additional conditions
(\ref{appendix/converse/addcond/Xu/Xu})--(\ref{appendix/converse/addcond/Xphi/Xphi}).
First,
from condition (\ref{appendix/converse/addcond/Xphi/Xphi}) we obtain
\begin{equation}
\begin{split}
	&\partial_{\varphi}{\vect{X}(u,\varphi)}\cdot\partial_{\varphi}{\vect{X}(u,\varphi)}%
	=%
		\stE^{2\stWeylScalingExp(u)}%
	\\
	&\quad%
	=%
		\tfrac{1}{2}%
		\stE^{2\stWeylScalingExp(u)}%
		\\
		&\cdot%
		\Bigl[%
			\bigl(%
				{\vect{A}}^{2}\!+\!{\vect{B}}^{2}%
			\bigr)%
			+%
			\cos{2\varphi}%
			\;%
			\bigl(%
				{\vect{A}}^{2}\!-\!{\vect{B}}^{2}%
			\bigr)%
			-%
			\sin{2\varphi}%
			\:%
			{\vect{A}}\!\cdot\!{\vect{B}}%
		\Bigr]%
	,
\end{split}
\end{equation}
or
\begin{equation}
	{\vect{A}}^{2}={\vect{B}}^{2}=1%
	\quad\text{and}\quad%
	{\vect{A}}\cdot{\vect{B}}=0%
	.
\end{equation}
Therefrom,
condition (\ref{appendix/converse/addcond/Xu/Xphi}) holds:
\begin{equation}
\begin{split}
	&\partial_{u}{\vect{X}(u,\varphi)}\cdot\partial_{\varphi}{\vect{X}(u,\varphi)}=%
		0%
	\\
	&\qquad%
	=%
		\stE^{2\stWeylScalingExp(u)}%
		\sin\stNPolarAngle(u)%
		\bigl[%
			\sin\varphi%
			\:%
			{\vect{A}}\!\cdot\!{\vect{C}}%
			-%
			\cos\varphi%
			\:%
			{\vect{B}}\!\cdot\!{\vect{C}}%
		\bigr]%
	,
\end{split}
\end{equation}
or
\begin{equation}
	{\vect{A}}\cdot{\vect{C}}=%
	{\vect{B}}\cdot{\vect{C}}=%
		0
	.
\end{equation}
Then, condition (\ref{appendix/converse/addcond/Xu/Xu}) reads
\begin{equation}
	\partial_{u}{\vect{X}(u,\varphi)}\cdot\partial_{u}{\vect{X}(u,\varphi)}=%
		\stE^{2\stWeylScalingExp(u)}%
	=%
		\stE^{2\stWeylScalingExp(u)}%
		\:%
		{\vect{C}}^{2}
	,
\end{equation}
or
\begin{equation}
	{\vect{C}}^{2}=1
	.
\end{equation}
In other words,
the three constant vectors of integration ${\vect{A}}$, ${\vect{B}}$, and ${\vect{C}}$
must form an orthonormal triplet of constant vectors.
Henceforth,
since no additional condition has to be fulfilled,
the constant vector of integration ${\vect{D}}$ is really an arbitrary constant vector,
namely an arbitrary translation.
In order to determine
the orientation of the orthonormal triplet,
we may explicitly compute
the outward surface unit normal vector $\widehat{\vect{N}}(u,\varphi)$:
\begin{multline}
\label{appendix/reconstruction/N/step/fourth/preliminary}
	\widehat{\vect{N}}(u,\varphi)=%
		\sin\stNPolarAngle(u)\cos\varphi\:%
			{\vect{B}}\times{\vect{C}}%
		\\
		+%
		\sin\stNPolarAngle(u)\sin\varphi\:%
			{\vect{C}}\times{\vect{A}}%
		+%
		\cos\stNPolarAngle(u)\:%
			{\vect{A}}\times{\vect{B}}%
	.
\end{multline}
By inserting
previous formula (\ref{appendix/reconstruction/N/step/fourth/preliminary})
and formula (\ref{appendix/reconstruction/X/step/third}),
Eq.~(\ref{appendix/converse/weingarten/phi}) holds:
\begin{equation}
	{\vect{C}}=%
		{\vect{A}}\times{\vect{B}}
	,
\end{equation}
thus the orthonormal triplet is direct.
To stress the nature of the direct orthonormal triplet
\mbox{$[{\vect{A}},{\vect{B}},{\vect{C}}]$}
and of the translational vector ${\vect{D}}$,
we may denote them by 
\mbox{$[{\widehat{\vect{e}}}_{\textsl{x}},{\widehat{\vect{e}}}_{\textsl{y}},{\widehat{\vect{e}}}_{\textsl{z}}]$}
and ${\vect{t}}$, respectively:
eventually
formula (\ref{appendix/reconstruction/X/step/third}) may read
\begin{multline}
\label{appendix/reconstruction/X/step/fourth}
	{\vect{X}(u,\varphi)}=%
		\stE^{\stWeylScalingExp(u)}%
		\bigl[%
			\cos\varphi%
				\:%
				{\widehat{\vect{e}}}_{\textsl{x}}%
			+%
			\sin\varphi%
				\:%
				{\widehat{\vect{e}}}_{\textsl{y}}%
		\bigr]%
		\\
		-%
		\int^{u}\!\!\!\!%
			\std{w}%
			\;%
			\stE^{\stWeylScalingExp(w)}%
			\sin\stNPolarAngle(w)%
			\:%
			{\widehat{\vect{e}}}_{\textsl{z}}%
		+%
			{\vect{t}}
	,
\end{multline}
which is the equation of an axisymmetric surface
revolving around the axis ${\widehat{\vect{e}}}_{\textsl{z}}$ shifted along ${\vect{t}}$
with radius $r(u)$ and height $z(u)$ given by
\begin{subequations}
\label{appendix/reconstruction/X/step/fifth}
\begin{align}
\label{appendix/reconstruction/X/step/fifth/radius}
	r(u)&=%
		\stE^{\stWeylScalingExp(u)}%
	,\\
\intertext{and}%
\label{appendix/reconstruction/X/step/fifth/height}
	z(u)&=%
		-%
		\int^{u}\!\!\!\!%
			\std{w}%
			\;%
			\stE^{\stWeylScalingExp(w)}%
			\sin\stNPolarAngle(w)%
	.
\end{align}
\end{subequations}
By the choice of the direct orthonormal triplet 
\mbox{$[{\widehat{\vect{e}}}_{\textsl{x}},{\widehat{\vect{e}}}_{\textsl{y}},{\widehat{\vect{e}}}_{\textsl{z}}]$}
and of the translation ${\vect{t}}$,
the surface of revolution (\ref{appendix/reconstruction/X/step/fourth})
can be placed in any position in space.
Finally let us notice that
the ultimate differential equation to integrate
(\ref{appendix/reconstruction/X/step/third/amble})
translates within this choice of notation into
\begin{equation}
\label{appendix/reconstruction/X/step/fifth/height/postamble}
	\partial_{u}{z(u)}=%
		-%
		\stE^{\stWeylScalingExp(u)}%
		\sin\stNPolarAngle(u)%
	.
\end{equation}

\subsection{Curvatures}
Whereas the intrinsic curvature
(or the Gaussian curvature)
$\stGC$ characterizes the metric tensor ${\tens{g}}_{ij}$
through the Riemann tensor ${\tens{R}}_{ijkl}$,
\begin{equation}
	\stGC=%
		\tfrac{1}{2}%
		\:%
		{\tens{g}}^{ij}%
		{\tens{g}}^{kl}%
		{\tens{R}}_{ikjl}
	,
\end{equation}
the mean curvature $\stMC$ and the extrinsic curvature $\stEC$
characterize the shape tensor ${\tens{b}}_{ij}$:
the mean curvature $\stMC$ is its half trace,
\begin{equation}
	\stMC=%
		\tfrac{1}{2}%
		\:%
		{\tens{g}}^{ij}%
		{\tens{b}}_{ij}%
	,
\end{equation}
while the extrinsic curvature $\stEC$ is its determinant,
\begin{equation}
	\stEC=%
		\tfrac{1}{2}%
		\:%
		\stLevCS^{ij}%
		\stLevCS^{kl}%
		{\tens{b}_{ik}}%
		{\tens{b}_{jl}}%
	.
\end{equation}
The tensor $\stLevCS_{ij}$ being the totally antisymmetric tensor:
\begin{equation}
	\stLevCS_{ij}=%
		\sqrt{\left|g\right|}\:%
		\stLevCSymb_{ij}%
		\quad\text{with}\quad%
		\stLevCSymb_{ij}=%
			\left[%
				\begin{smallmatrix}
					\hphantom{-}0&+1\\
					-1&\hphantom{+}0
				\end{smallmatrix}
			\right]%
	.
\end{equation}
Therefore,
in isometric azimuthal coordinates \mbox{$(u,\varphi)$},
the intrinsic curvature $\stGC$ holds:
\begin{equation}
\label{appendix/curvature/intrinsic}
	\stGC(u)=%
		-%
		\stE^{-2\stWeylScalingExp(u)}%
		\:%
		\partial_{uu}\stWeylScalingExp(u)%
	,
\end{equation}
according to formula (\ref{appendix/tensor/riemann}).
On the other hand,
the mean curvature $\stMC$ yields
\begin{equation}
\label{appendix/curvature/mean}
	\stMC(u)=%
		-\tfrac{1}{2}%
		\:%
		\stE^{-\stWeylScalingExp(u)}%
		\left[%
			{\partial_{u}\stNPolarAngle(u)}%
			+%
			{\sin\stNPolarAngle(u)}%
		\right]%
	,
\end{equation}
and the extrinsic curvature $\stEC$ verifies
\begin{equation}
	\stEC(u)=%
		\stE^{-2\stWeylScalingExp(u)}%
		\:%
		{\sin\stNPolarAngle(u)}%
		{\partial_{u}\stNPolarAngle(u)}%
	.
\end{equation}
By invoking the elastic compatibility condition (\ref{appendix/intcond/codazzi}),
it is easily checked that
the intrinsic curvature $\stGC$ and the extrinsic curvature $\stEC$
are effectively the same as asserted by the Gauss theorem.

\section{Recovering the Shape Equation in the Absence of ``Artifact"}
\label{appendix/B}
The customary generic shape equation corresponding to our system
without the ``artifact" is the following covariant equation \cite{SEASVC}:
\begin{equation}
\label{appendix/GSE/literature/compact}
	\tfrac{1}{2}\triangle\stMC%
	+%
	\stMC%
	\bigl[%
		\stMC^{2}-\stGC%
	\bigr]%
	=%
		0
	,
\end{equation}
where $\triangle$ denotes the Laplacian,
$\stMC$ the mean curvature,
and $\stGC$ the extrinsic curvature.
Once the Laplacian is expanded,
the generic shape equation (\ref{appendix/GSE/literature/compact}) writes
\begin{equation}
	\tfrac{1}{2}%
	\bigl(%
		{\tens{g}}^{il}\partial_{l}%
		-%
		{\tens{g}}^{mn}\Gamma^{i}_{\hphantom{x}{mn}}
	\bigr)%
	\partial_{i}\stMC
	+%
	\stMC%
	\bigl[%
		\stMC^{2}-\stGC%
	\bigr]%
	=%
		0%
	,
\end{equation}
or, in isometric azimuthal coordinates \mbox{$(u,\varphi)$},
\begin{equation}
\label{appendix/GSE/literature/expanded}
	\tfrac{1}{2}\:%
	\stE^{-2\stWeylScalingExp(u)}%
	\partial_{uu}\stMC(u)%
	+%
	\stMC(u)%
	\bigl[%
		\stMC^{2}(u)-\stGC(u)%
	\bigr]%
	=%
		0%
	.
\end{equation}
Substituting formulas (\ref{appendix/curvature/intrinsic}) and 
(\ref{appendix/curvature/mean}) into previous
Eq.~(\ref{appendix/GSE/literature/expanded}) leads to
\begin{multline}
\label{appendix/GSE/literature/coord/ia}
	\bigr(%
		{\partial_{u}}%
		-%
		{\cos\stNPolarAngle(u)}
	\bigl)%
	\bigr(%
		{\partial_{uu}\stNPolarAngle(u)}%
		-%
		{\sin\stNPolarAngle(u)}%
		\,%
		{\cos\stNPolarAngle(u)}%
	\bigl)%
	\\
	+%
	\tfrac{1}{2}%
	\bigl(%
		{\partial_{u}\stNPolarAngle(u)}
		-%
		{\sin\stNPolarAngle(u)}%
	\bigr)^{2}%
	\bigl(%
		{\partial_{u}\stNPolarAngle(u)}
		+%
		{\sin\stNPolarAngle(u)}%
	\bigr)%
	=%
		0%
	.
\end{multline}

On the other hand,
our generic shape equation (\ref{BP/SVR/Fb/functional/Beltrami/generic})
without the ``artifact" reads
\begin{multline}
\label{appendix/appendix/GSE/Beltrami/coord/ia}
	-\tfrac{1}{2}%
		\bigl(%
			{\partial_{u}\stNPolarAngle(u)}%
		\bigr)^{2}%
	+%
	\tfrac{1}{2}%
		{\sin^{2}\stNPolarAngle(u)}%
	+%
	\stLmCI\:%
		{\cos\stNPolarAngle(u)}%
	\\
	+%
	\stLnCI%
		\bigl[%
			z(u)\,%
			{\cos\stNPolarAngle(u)}%
			+%
			r(u)\,%
			{\sin\stNPolarAngle(u)}%
		\bigr]%
	=%
		0%
	.
\end{multline}
In order to compare the shape equation (\ref{appendix/appendix/GSE/Beltrami/coord/ia})
with the general shape equation (\ref{appendix/GSE/literature/coord/ia}),
let us first transform it in a form free of
radius $r(u)$, height $z(u)$, and constants of integration $\stLmCI$ and $\stLnCI$
as follows.
Since formulas (\ref{appendix/reconstruction/X/step/fifth/radius}),
(\ref{appendix/intcond/codazzi}),
and (\ref{appendix/reconstruction/X/step/fifth/height/postamble})
yield the equality
\begin{equation}
	{\partial_{u}{z(u)}}\,%
	{\cos\stNPolarAngle(u)}%
	+%
	{\partial_{u}{r(u)}}\,%
	{\sin\stNPolarAngle(u)}%
	=%
		0
	,
\end{equation}
the differentiation of Eq.~(\ref{appendix/appendix/GSE/Beltrami/coord/ia})
readily gives
\begin{multline}
\label{appendix/appendix/GSE/Beltrami/coord/ia/diff}
	{\partial_{uu}\stNPolarAngle(u)}%
	-%
	{\sin\stNPolarAngle(u)}%
	\,%
	{\cos\stNPolarAngle(u)}%
	=%
		-\stLmCI\:%
		{\sin\stNPolarAngle(u)}%
		\\
		+%
		\stLnCI%
			\bigl[%
				-%
				{z(u)}\,%
				{\sin\stNPolarAngle(u)}%
				+%
				{r(u)}\,%
				{\cos\stNPolarAngle(u)}%
			\bigr]%
	.
\end{multline}
By applying the operator
$%
	\bigr(%
		{\partial_{u}}%
		-%
		{\cos\stNPolarAngle(u)}
	\bigl)%
$
to both sides of previous equality
(\ref{appendix/appendix/GSE/Beltrami/coord/ia/diff}),
we obtain
\begin{multline}
\label{appendix/appendix/GSE/Beltrami/coord/ia/ddiff}
	\bigr(%
		{\partial_{u}}%
		-%
		{\cos\stNPolarAngle(u)}
	\bigl)%
	\bigr(%
		{\partial_{uu}\stNPolarAngle(u)}%
		-%
		{\sin\stNPolarAngle(u)}%
		\,%
		{\cos\stNPolarAngle(u)}
	\bigl)%
	=%
	\\
		-%
		\bigr(%
			\stLmCI\:%
				{\cos\stNPolarAngle(u)}%
			+%
			\stLnCI%
				\bigl[%
					z(u)\,%
					{\cos\stNPolarAngle(u)}%
					+%
					r(u)\,%
					{\sin\stNPolarAngle(u)}%
				\bigr]%
		\bigl)%
		\\
		\cdot%
		\bigr(%
			{\partial_{u}\stNPolarAngle(u)}%
			-%
			{\sin\stNPolarAngle(u)}%
		\bigl)%
	,
	\qquad
\end{multline}
after simplification of the right-hand side.
Then the insertion of formula (\ref{appendix/appendix/GSE/Beltrami/coord/ia})
into previous Eq.~(\ref{appendix/appendix/GSE/Beltrami/coord/ia/ddiff})
gives Eq.~(\ref{appendix/GSE/literature/coord/ia}).

In summary,
as announced,
the two shape equations (\ref{appendix/appendix/GSE/Beltrami/coord/ia})
and (\ref{appendix/GSE/literature/coord/ia}) are exactly the same.

\bibliographystyle{apsrev}
\bibliography{svdglbes}

\begin{thebibliography}{44}
\bibitem[{Ein()}]{EinsteinSummationConvention}
\bibinfo{note}{Repeated indices are summed over --- indices being lowered and
  raised with ${\tens{g}}_{ij}$ and its inverse ${\tens{g}}^{ij}$,
  respectively.}
\end{thebibliography}

\end{document}